\pdfoutput=1 
\documentclass{JINST}
\let\ifpdf\relax

\usepackage{siunitx}
\usepackage{pdflscape}
\usepackage{gensymb}
\usepackage{graphicx}

\title{Scalability, scintillation readout and charge drift in a kilogram scale solid xenon particle detector}

\author{J.~Yoo\thanks{Corresponding Author: yoo@fnal.gov}, H.~Cease, W.~F.~Jaskierny, D.~Markley, and R.~B.~Pahlka\\
Fermi National Accelerator Laboratory, Kirk and Pine St., Batavia, IL 60510, USA}
\author{D.~Balakishiyeva and T.~Saab\\
Department of Physics, University of Florida, Gainesville, FL 32611, USA}
\author{M.~Filipenko\\
Erlangen Center for Astroparticle Physics (ECAP), Friedrich Alexander
University of Erlangen-Nuremberg, Erwin-Rommel-Stra\"sse 1, 91058
Erlangen, Germany\\
}

\abstract{We report a demonstration of the scalability of optically transparent xenon in the solid phase for use as a particle detector above a kilogram scale. We employ a liquid nitrogen cooled cryostat combined with a xenon purification and chiller system to measure the scintillation light output and electron drift speed from both the solid and liquid phases of xenon. Scintillation light output from sealed radioactive sources is measured by a set of high quantum efficiency photomultiplier tubes suitable for cryogenic applications. We observed a reduced amount of photons in solid phase compared to that in liquid phase. We used a conventional time projection chamber system to measure the electron drift time in a kilogram of solid xenon and observed faster electron drift speed in the solid phase xenon compared to that in the liquid phase.}

\preprint{FERMILAB-PUB-14-402-E}
\maketitle

\begin{document}
\section{Introduction}\label{intro}
\par The Standard Model (SM) of Particle Physics has been explored with amazing accuracies from the scale of the Hubble radius to the size of nucleons.  Despite the remarkable success of the SM, often regarded as a triumph of modern physics, a number of observations have recently emerged suggesting the incompleteness of our understanding of fundamental interactions. Three main pieces of evidence contribute to this conclusion: the CP-asymmetry of the Universe, the existence of dark matter and the discovery of massive neutrinos. In all these cases, low background experiments in deep underground sites provide excellent venues to discover new physics Beyond the Standard Model. 

\par Noble elements in both the gas and liquid phases have proven use as excellent low background radiation detectors~\cite{aprile2010,ackerman2011,aprile2012,akerib2014,kamland2013}. Xenon has drawn special attention among the noble elements due to several distinct advantages over its lighter counterparts. The liquid phase of xenon possesses a very high scintillation light yield (40$\sim$60\, photons/keV) and the vacuum ultra-violet (VUV) wavelength (178\,nm)~\cite{jortner1965} of the scintillation is optically transparent in xenon~\cite{Szydagis:2011tk}. Xenon also has good ionization and electron transport properties; the absence of long-lived radioisotopes in xenon results in no intrinsic background radiation sources. The large atomic mass of xenon (Z=54) keeps external radioactive decay sources at the outer surfaces of the condensed xenon detector, resulting in improved self-shielding effects. As xenon is a noble element, the chemical purification is straightforward using hot getter and/or gas distillation systems that can remove most of the non-noble contaminants. A wide scope of applications has been studied on particle tracking and spectroscopy including $\gamma$-ray astronomy, neutrinoless double beta decay, dark matter searches, neutrino coherent scattering experiments and medical imaging devices.

\par The solid (crystalline) phase of xenon inherits most of these advantages from liquid xenon as a particle detector material. Electron drift speeds in thin layers of solid xenon are measured to be faster compared to those in the liquid phase~\cite{Miller:1968zza} and an increased amount of light collection in solid compared to liquid per given energy deposit in alpha irradiation has been reported~\cite{Aprile1994129}. Xenon in the solid phase is chemically stable and forms simple face centered cubic crystal structures with interatomic binding energies coming from weak Van der Waals forces. It is known to be relatively easy to produce small scale clear, transparent solid noble element specimens which are virtually perfect crystals but nevertheless are polycrystalline and contain a large number of microscopic defects~\cite{RGSv2}. The crystal structure~\cite{Venables1972, Kramer1972, Kramer1976, Niebel1974}, microscopic defects~\cite{Venables1966}, and directionality are important properties for certain experimental applications~\cite{Ahmed:2009ht}. However, for most particle detector applications, it is important to first understand the macroscopic properties such as density uniformity, optical transparency, charge drift velocity, scintillation, and ionization. Particle detectors based on solid phase noble element active media have been investigated thoroughly~\cite{Bolozdynya1977, Himi1982, Kubota1982, Kink1987, Bald1962,  Varding1994, Aprile1994129, Michniak2002387, Gushchin1982, Miller:1968zza}. Even though most of these studies have successfully shown that the solid noble elements are excellent candidates for particle detector material, large scale detectors have yet to be realized. Our R\&D effort, therefore, is focused on particle detector application of solid xenon with three practical goals: (1) proof of scaleability of optically transparent solid xenon which is distinguishable from opaque frozen xenon bulk, (2) measuring the scintillation light properties in solid xenon under radiation interactions using VUV Photo Multiplier Tubes (PMTs), and (3) measuring the electron drift time in large scale solid xenon using a conventional time projection chamber (TPC). 

\par In this paper, we describe the instrument details, demonstrate the scalability of solid xenon, and assess the scintillation readout and charge drift in a kg scale of xenon in the solid phase. We describe the cryogenic instruments in Section~\ref{sec:teststand}. The scalability of optically transparent solid xenon and methods are explained in Section~\ref{sec:scalability}. The scintillation readout and electron drift are discussed in Sections~\ref{sec:scintillation}, and~\ref{sec:chargetransport} respectively. We discuss the results in Section~\ref{sec:discussion} and summarize this paper in Section~\ref{sec:summary}.

\section{Solid xenon test stand}\label{sec:teststand}
\par
\begin{figure}[!t]
\begin{center}
	\includegraphics[width=1\columnwidth]{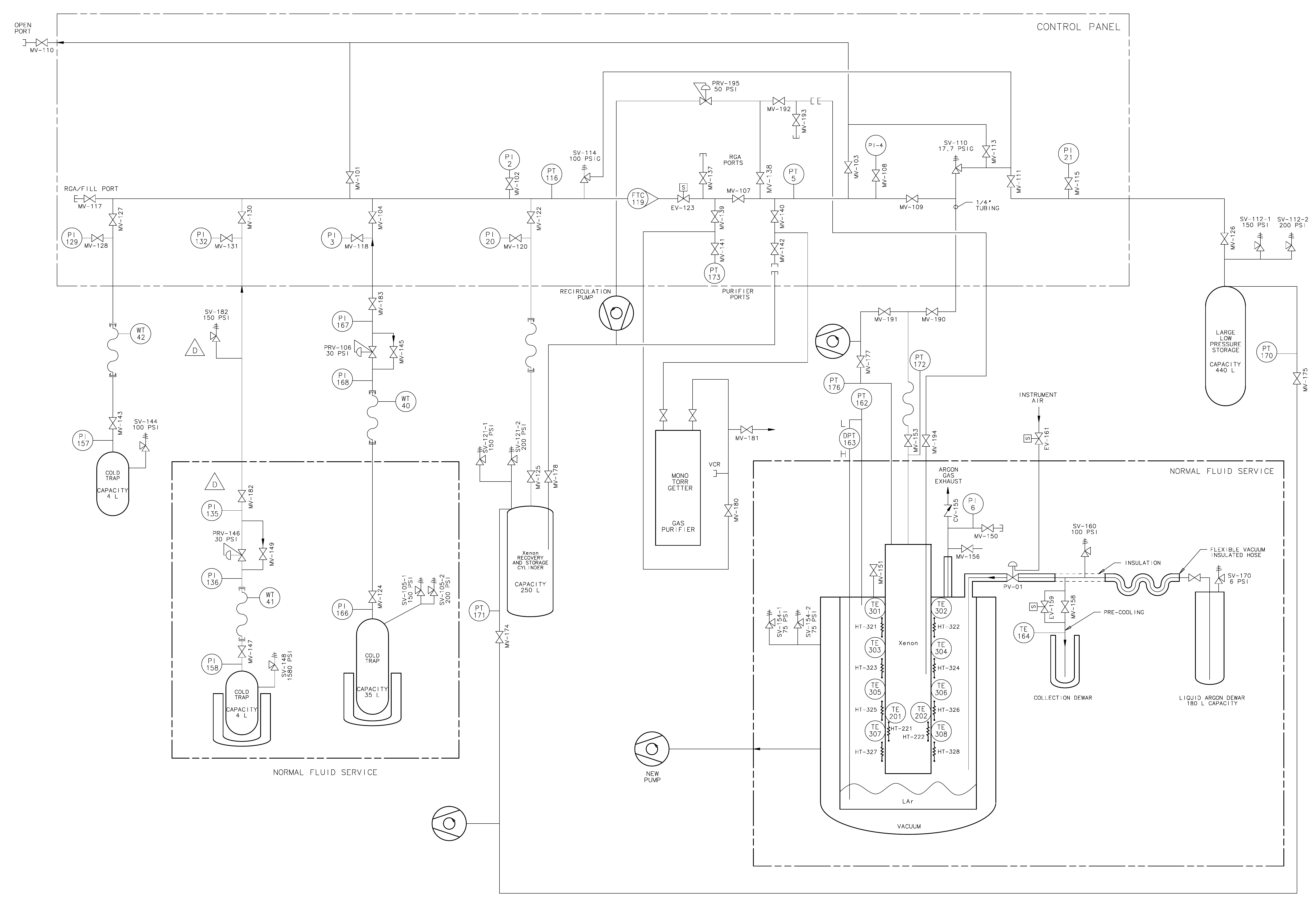}
	\caption{A piping diagram of the solid xenon test stand~\label{fig:pnid}}
\end{center}
\end{figure}

 The solid xenon test stand is located at the Proton Assembly Building at Fermi National Accelerator Laboratory.  It includes a principal xenon cryostat, a 440 liter xenon recovery cylinder, a 250 liter buffer tank, two 4 liter xenon storage cylinders and a 4 liter cylinder for calibration gas. A commercial hot getter (PF4-C3-R-1 and Monotorr PS4-MT3-R-1 by SAES) and a circulation loop allows continuous purification of the xenon.  Figure~\ref{fig:pnid} shows the piping diagram of the solid xenon test stand. 

\par The cryostat consists of a stainless steel vacuum jacketed chamber with an outermost diameter of 30\,cm, which has three 15\,cm diameter glass window ports. Two concentrically placed glass chambers reside inside allowing optical access to the xenon bulk volume.  The larger of the two glass chambers is used as a liquid nitrogen bath for cooling and has a diameter of 23\,cm, referred to as the LN (liquid nitrogen) chamber. The smaller 10\,cm diameter inner chamber houses the xenon volume and is made out of Pyrex with a 5\,mm thick side wall, and a 10\,mm thick flat bottom, referred to as the xenon chamber. 

\begin{figure}[t]
\begin{center}
\includegraphics[height=0.6\columnwidth]{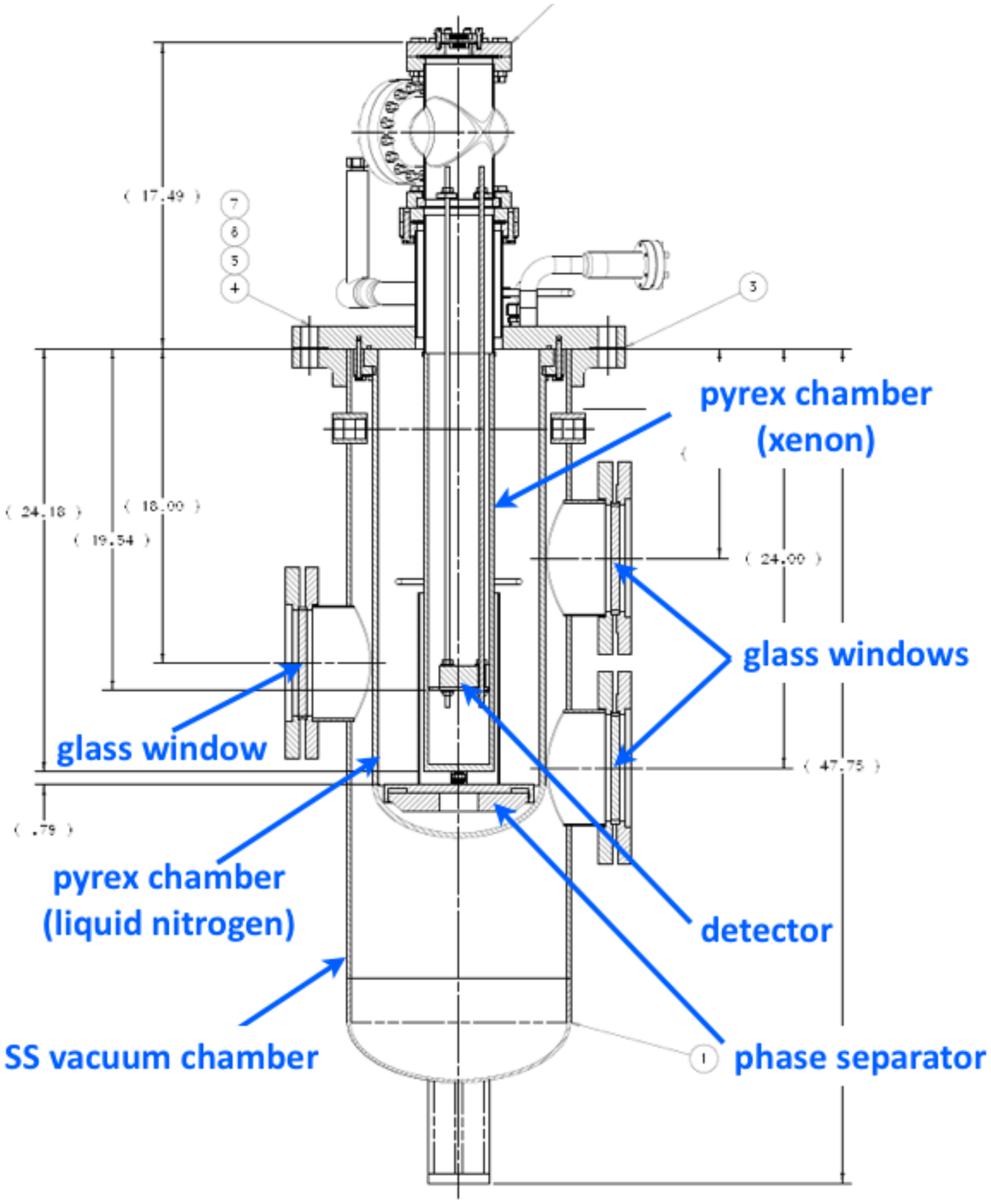}
\includegraphics[height=0.6\columnwidth]{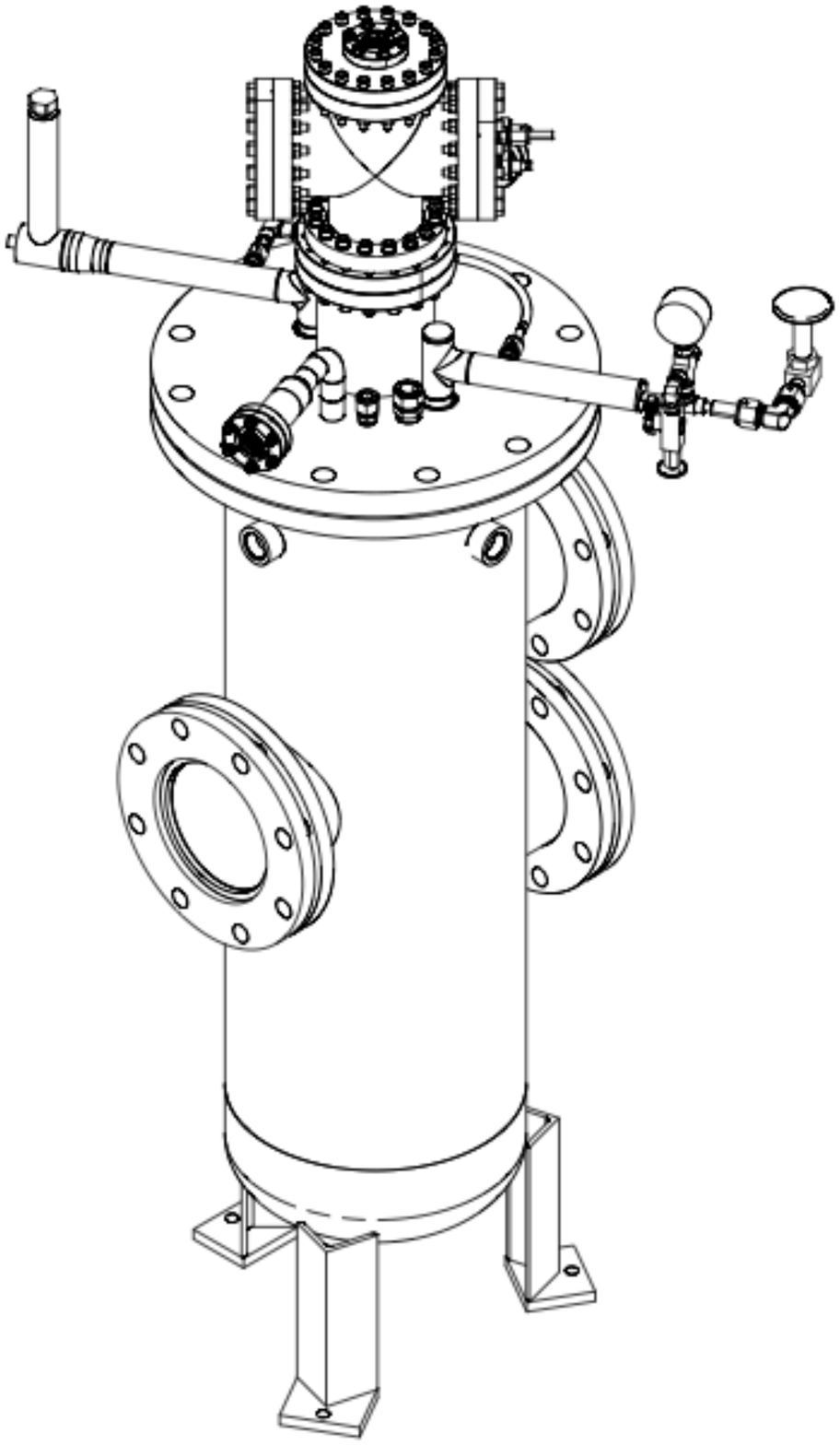}
\caption{Section view of the solid xenon chamber system and isometric view of the external chamber. \label{fig:sxchamber}}
\end{center}
\end{figure}

Figure~\ref{fig:sxchamber} shows schematics of the solid xenon cryostat and its isometric view. The stainless steel chamber also functions as a safety protection chamber in case of unexpected pressure changes in the glass chambers. The LN chamber is equipped with an effective phase separator at the bottom, made with a combination of aluminum and polyethylene blocks. The phase separator, when it is pre-cooled to liquid nitrogen temperature, transfers nitrogen liquid into the LN chamber with minimal evaporation in the internal transfer line.  

\par We employ ten Nichrome 32 gauge heater wires and Platinum Resistance Temperature Detectors (RTD: PT-100) that are affixed at the bottom and barrel surface of the xenon chamber using cryogenic epoxy. A total of nine RTDs measure the thermal gradients of the outer surface of the xenon chamber and one thermometer-heater loop is installed near the bottom of the inner wall of the xenon chamber.  While the cryogenic features of the system reported here largely overlap with those of liquid-based systems, we require additional advanced controls for temperature and pressure in order to properly solidify the xenon. The glass chambers place constraints on the pressure in the inner glass chamber, set to about 1\,bar.  The xenon chamber is manufactured with a allowable tensile stress of 7250\,PSI, a Young's modulus of 10.5$ \times 10^6$\,PSI, and a Poisson's ratio of 0.16. Reducing the yield strength by a factor of ten gives a maximum allowable stress of 725\,PSI. We carried out a simulation of
mechanical durability against pressurization of the glass chamber. In order to allow a large safety factor margin, the pressure relief point of the xenon chamber is set to 17.5\,PSID. The stress in the glass is estimated over the maximum differential pressure range. Both hoop stress (170\,PSI) and longitudinal stress (85\,PSI) due to the differential pressure of 17.5\,PSI is much lower than the design stress 680\,PSI. In the case of unexpected xenon chamber pressurization, a safety valve will relieve the xenon gas to the 440 liter backup chamber which is pre-evacuated under $10^{-6}$\,Torr.

\par A main control panel is designed to handle the xenon transfer and monitoring at a centralized location. It allows convenient monitoring and controlling of setting parameters, such as the xenon flow rate, cooling bath level, temperatures, pressures, and weights of the xenon storage cylinders. A Programmable Logic Controller (PLC; Beckhoff CX1000) is used with KL I/O modules attached to the CX1000 K bus backplane. The slow control system has approximately twenty electronic input sensing devices and twenty output devices. Input devices include RTDs, pressure transmitters, liquid nitrogen level transmitters, and weight strain gauges. Output devices include solenoid valves, gas flow controllers, and heaters. The cryogenic control system has been designed and built following all the required standards including the National Electric Code. The local HMI (Human Machine Interface) is a C-More touch screen panel manufactured by Automation Direct. An HMI system by GEFANUC with a product name of iFIX system is used for remote controlling and historic data logging of the PLC. Ten individually controlled heaters are powered by 24\,VDC. The power to each heater is controlled by a DC solid state relay with a pulse control module firing the relay at a rate of 1\,Hz. 

\par The cooling LN is fed from a continuously-serviced 20 ton external LN tank that facilitates uninterrupted cryogenic system operation. Using the cold nitrogen gas above the liquid to cool the xenon chamber, the liquid nitrogen level in the LN chamber is controlled using a feed-back system of pneumatic control valves and differential pressure level meters with an accuracy of about 0.5\,cm. The liquid nitrogen level is set to about 2\,cm below the bottom of the inner glass chamber during normal operation and is therefore not in direct contact with the xenon chamber. The temperature of the glass chamber is controlled using three sets of thermometer--heater feedback loops (bottom, barrel, and top). The insulation vacuum between the stainless steel chamber and the LN chamber is maintained at about $10^{-5}$\,Torr using a dedicated turbo pump. 

\par The overall leakage level of the entire xenon gas handling system is measured to be better than $10^{-10}$\,Torr of helium gas leakage. The xenon chamber is cleaned in a ultrasonic bath before the installation and is then baked at 40\,C\degree ~for a few days using the attached heaters. The vacuum level of the xenon vessel reaches $10^{-8}$\,Torr at room temperature with no detector instruments in the chamber and stabilizes at this value at a temperature of 164\,K (or below).  With detector instruments installed in the chamber, the vacuum level reaches below $10^{-5}$\,Torr after a couple of days of evacuation.

\section{Scalability of solid xenon}\label{sec:scalability}
\par The main focus of the scalability study is to understand the conditions required to produce optically transparent solid-phase xenon which is distinguishable from frozen opaque volumes or other types of solid phases. The microscopic defects, directionality and structure of the crystal are not the subjects of this study. Due to the the density difference between the liquid phase (2.95 g/cc) and the solid phase (3.41 g/cc), the growing process of solid xenon requires special care in maintaining and controlling the growing speed.  
 
\par The thermal condition within the chamber is designed to produce about 2 kg of solid xenon. We found that a transparent layer of solid xenon can be reliably grown via vapor deposition methods in a pre-cooled chamber. However, the growth rate of solid xenon is quite slow ($\sim$5\,mm/day initially, then gradually slows down), and the shapes of the solid surfaces were non-uniform. We concluded that the vapor deposition method would be more appropriate for thin layer detectors rather than large scale detectors. 

\par For a kg-scale (or more than a few centimeter thick) solid xenon test we adopted a modified {\it Bridgeman's technique}~\cite{RGSv2}. In Bridgeman's method, a temperature gradient of 1$\sim$2\,K/cm is established across the liquid, then growing from the liquid to solid at near 1\,bar of pressure, which is then cooled progressively from the bottom at a rate of $\sim$1 K/hr to allow the heat of solidification to dissipate, and for annealing to take place.

\begin{figure}[t!]
\begin{center}
\includegraphics[width=5in]{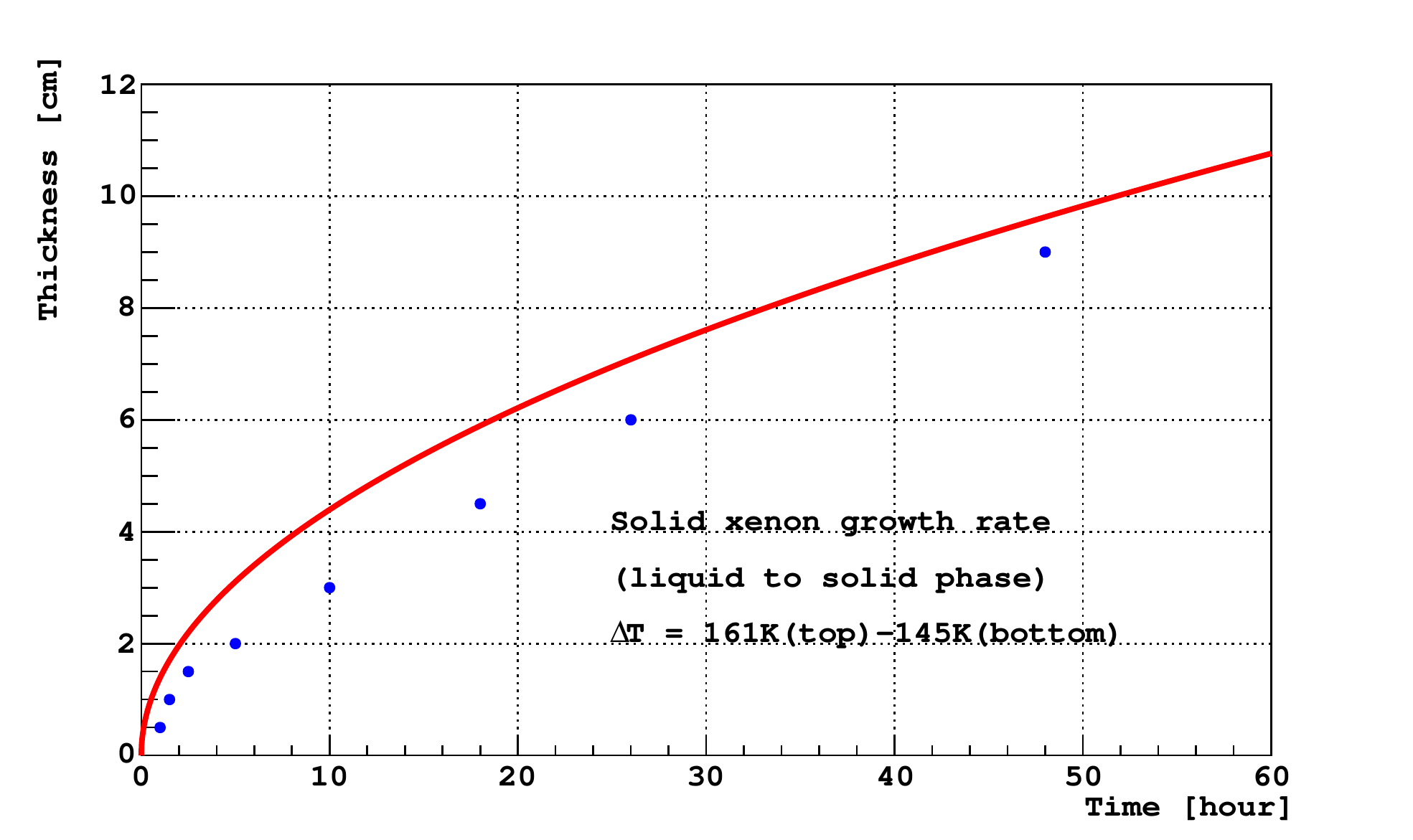}
\caption{A simple thermal model of the expected solid xenon growth rate (red curve), and an example of actual growth rate of solid xenon (blue dots). The bottom temperature of the liquid xenon volume is set to 145\,K and top temperature set to 161\,K.~\label{fig:sxgrow}}
\end{center}
\end{figure}

\begin{figure}[t!]
\begin{center}
\includegraphics[height=2.in]{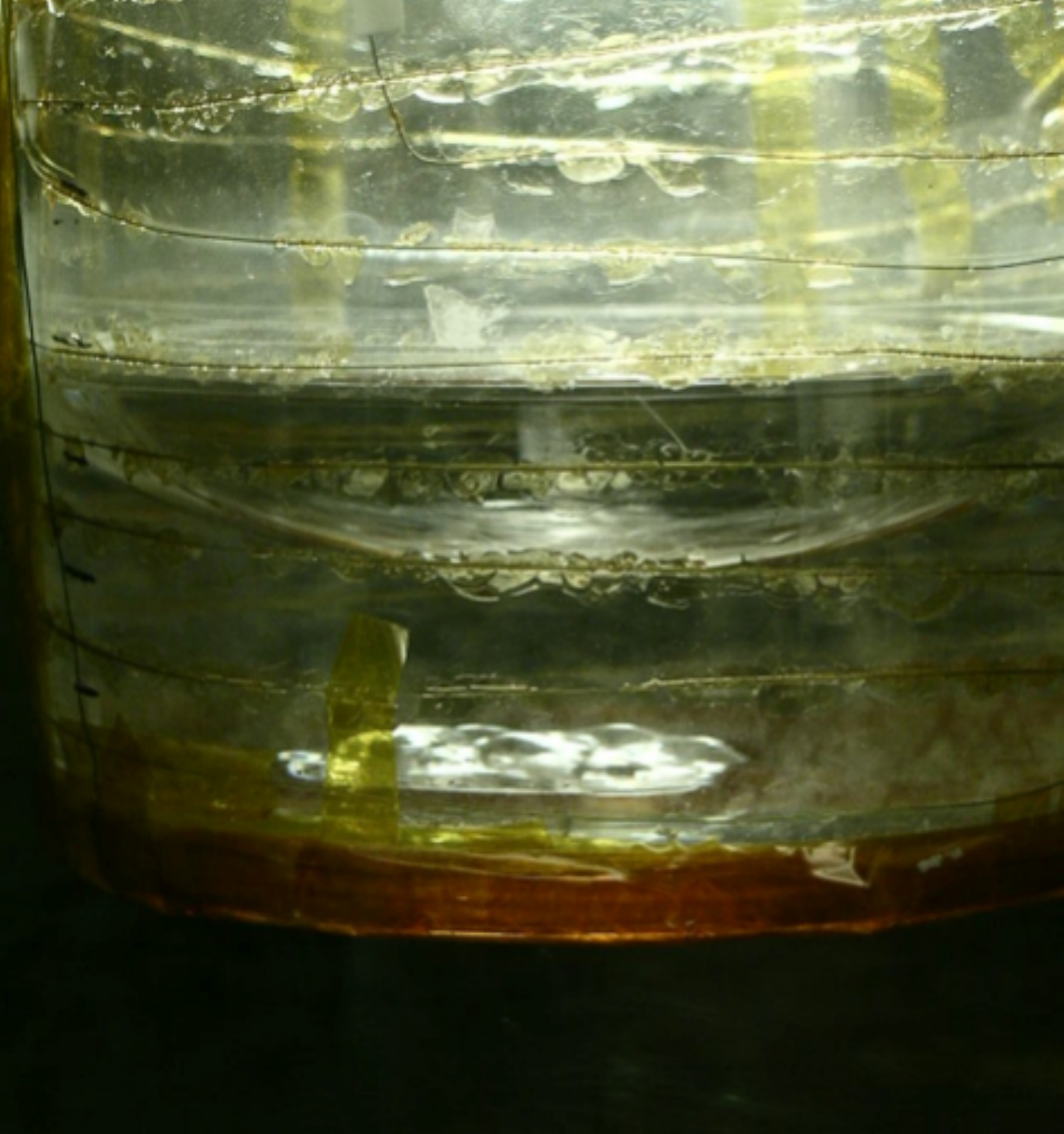}
\includegraphics[height=2.in]{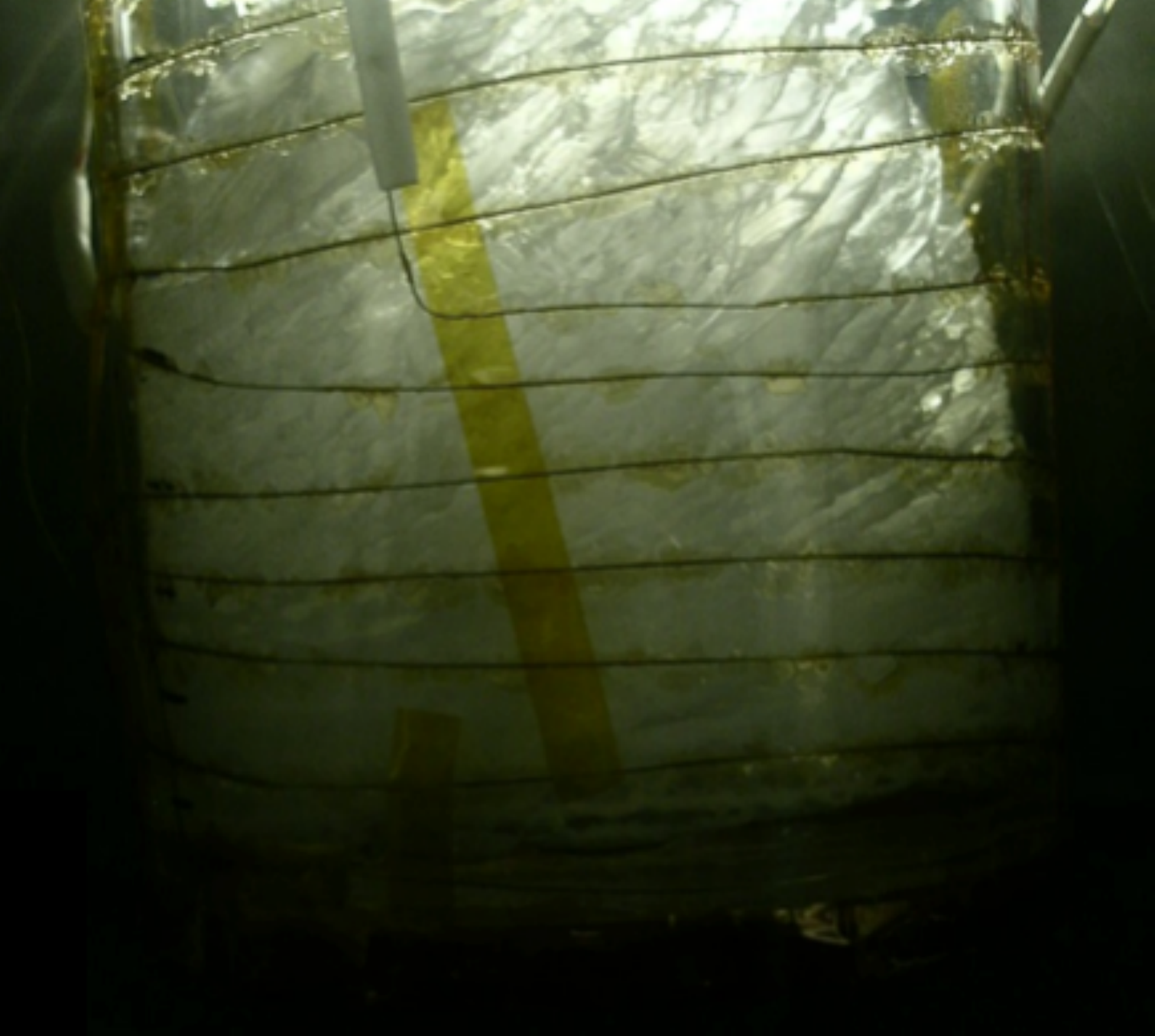}
\includegraphics[width=4.15in]{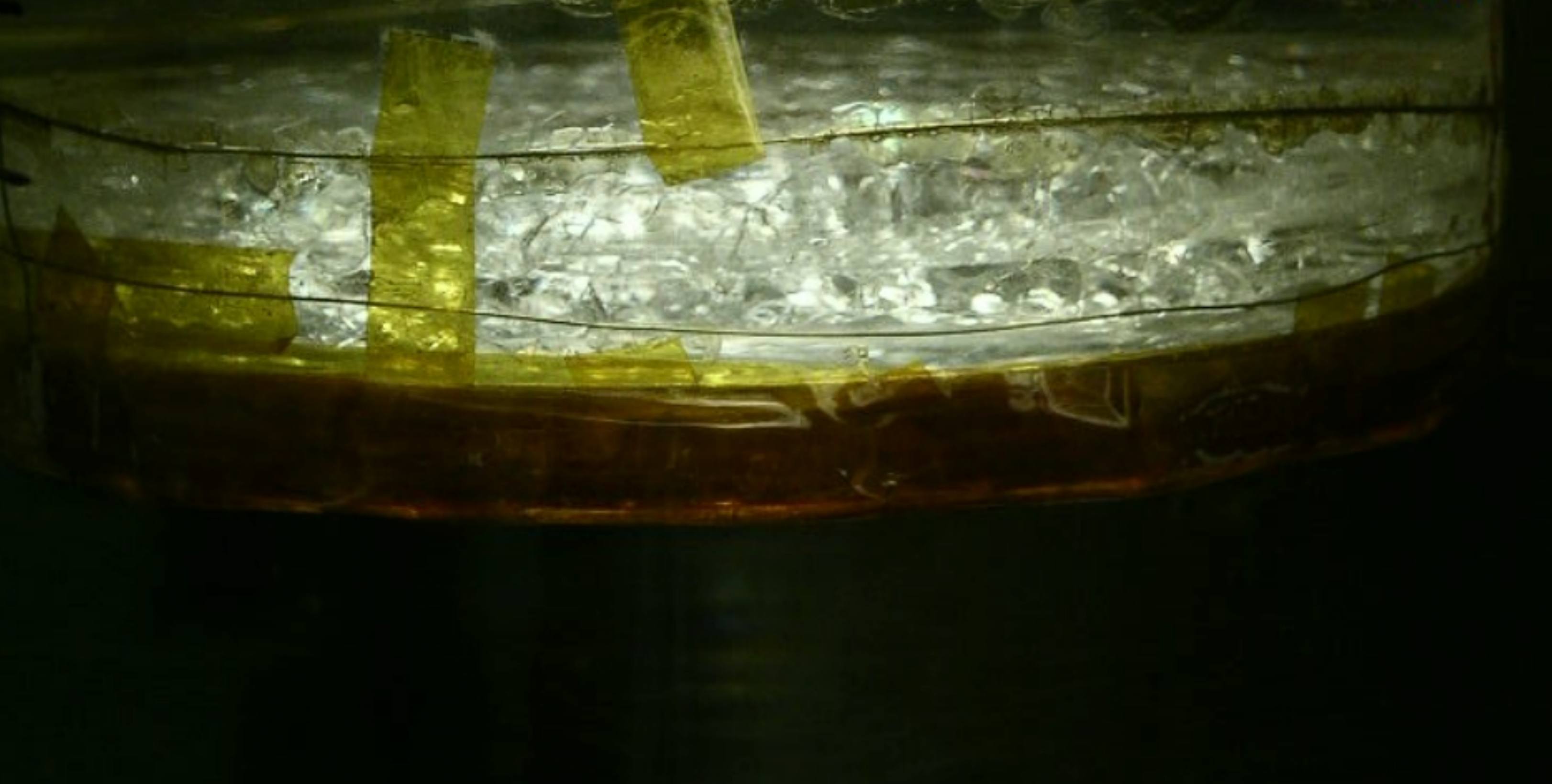}
\caption{Photographs of various types of solid xenon. The top left photo shows a concave structure of the solid surface due to the over-cooling at the barrel. There is also an opaque substructure at the bottom center due to the significant temperature gradient at the bottom surface. These are produced when there was a rapid variation of cooling liquid level. The top right photo shows frozen layers of solid xenon. These structures can be easily created when the pressure drops below the vapor pressure. The bottom photo shows solid xenon in a polycrystalline state on the order of a few mm, provided when the temperature and pressure control become unstable near the triple point (161.4\,K and 11.9 PSIA). The dark brown color at the bottom of the chamber is kapton tape. The spacing of the heater between each wire is about 1\,cm, which is a good measure of approximate height of xenon.~\label{fig:frozenxenon}}
\end{center}
\end{figure}

\par Due to the slow growth rate of the transparent solid phase of xenon, the bottom to top temperature gradient needs to be tuned depending on the size of the solid xenon. In ideal conditions, the growth rate of solid xenon from the liquid phase (thickness $L$, as a function of time $t$) can be estimated by $L = \sqrt{(2K~\Delta T)/(H\rho)} \sqrt{t}$, where the latent heat of xenon fusion is $H$=17.5\,J/g, the thermal conductivity of solid xenon is $K$=0.001\,W/(cm$\cdot$K)~\cite{Purskii2004}, and the solid xenon density is $\rho$=3.41/cc. Figure~\ref{fig:sxgrow} shows a simple thermal model of the expected solid xenon growth rate (red curve) and an example of actual growth data (blue dots). In this example, the bottom temperature is set to 145\,K and top temperature is set to 161\,K (liquid xenon temperature). The actual growth rate of solid xenon is a little slower than expected by the simple model. In order to grow 2\,kg (or about 9\,cm high in the xenon chamber with an area of 3.14$\times$(4.5\,cm)$^2$) of solid xenon, it requires more than 48 hours of stable temperature and pressure control. In reality, the details of the thermal configuration in the solid xenon chamber system requires fine tuning of the temperature gradient. We found the following setup reliably reproduces the optically transparent solid xenon in our system. 

\par First, the xenon is condensed and liquified in the glass chamber at a set temperature of 163\,K and pressure of 14.5$\pm$0.5\,PSIA. When the liquid xenon level reached about a kg (about 4.5\,cm height in the xenon chamber), the bottom of the chamber was slowly cooled down to 145$\pm$0.5\,K, while cooling the barrel part of the chamber to 157$\pm$0.5\,K. The thick glass bottom (1\,cm) and barrel wall (0.5\,cm) reduce the thermal shock at the inner surface of the glass. Once these temperatures are set, the solid xenon forms at the bottom of the liquid xenon with an initial growing rate of about a cm per hour and gradually slowed down. Under these conditions, the solid xenon grows almost uniformly over the 9\,cm diameter of surface area. We found the temperature balance at the barrel and bottom of the glass chamber is important for uniform growth of the solid surface. If the barrel temperature is too low (150\,K or lower), the growing speed at the barrel exceeds that of the center and begins to form a concave solid surface. Any substantial temperature change produces optical defects in the solid xenon; such as opaque spots, filamentary structures and voids. We do not observe any significant pressure dependence of the optical quality of the solid xenon so long as the pressure is kept well above the vapor pressure. Rapid pressure variations near the vapor pressure can easily produce opaque layers at the top surface of the solid volume. The failures of temperature and pressure stability control near the triple point creates a few millimeters of polycrystalline xenon which is clearly identified by the birefringence of each polycrystalline cell. Figure~\ref{fig:frozenxenon} shows a few examples of solid phases of xenon which were produced during the initial test of the solid xenon cryogenics.

\begin{figure}[t!]
\begin{center}
\includegraphics[height=2.3in]{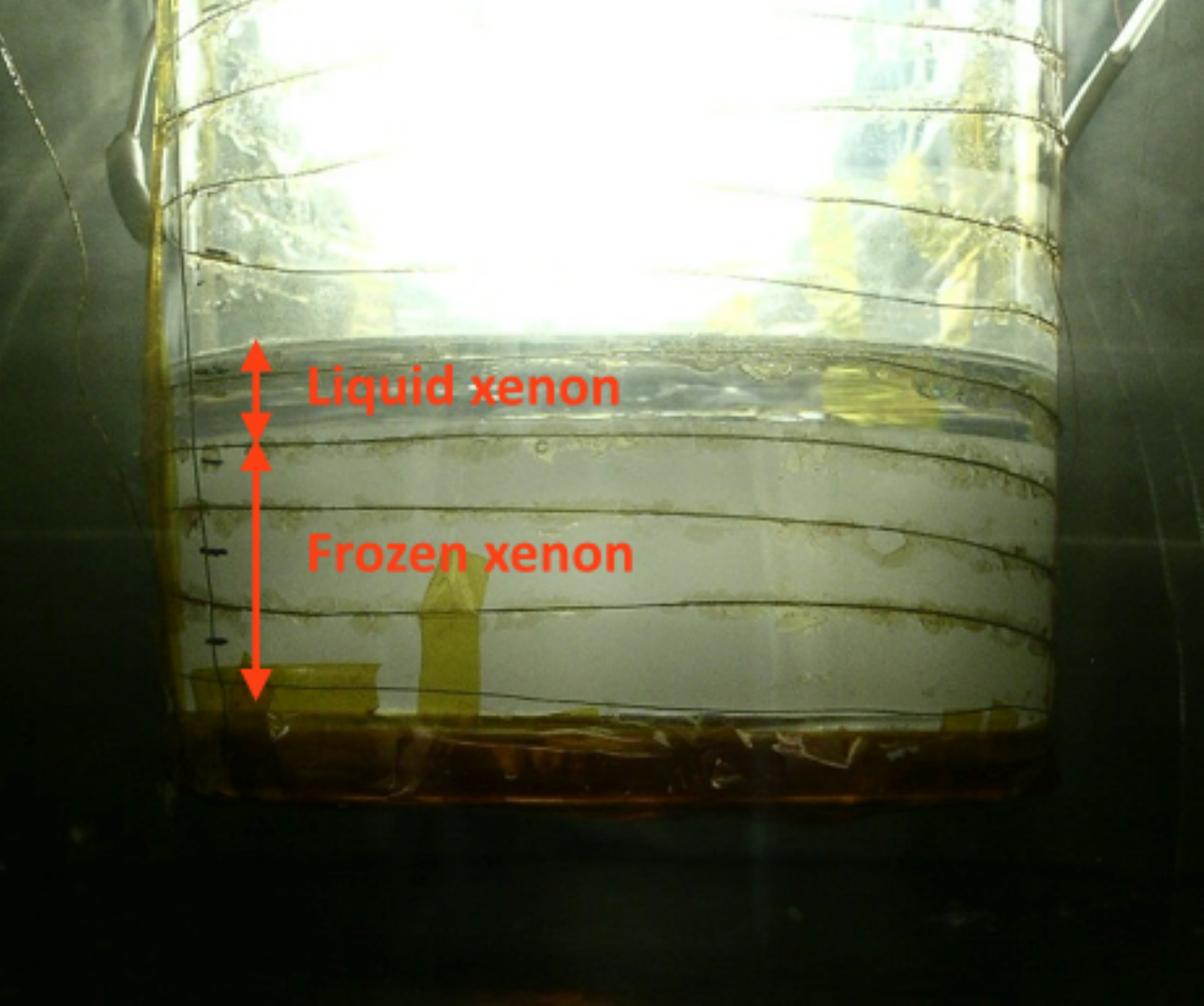}
\includegraphics[height=2.3in]{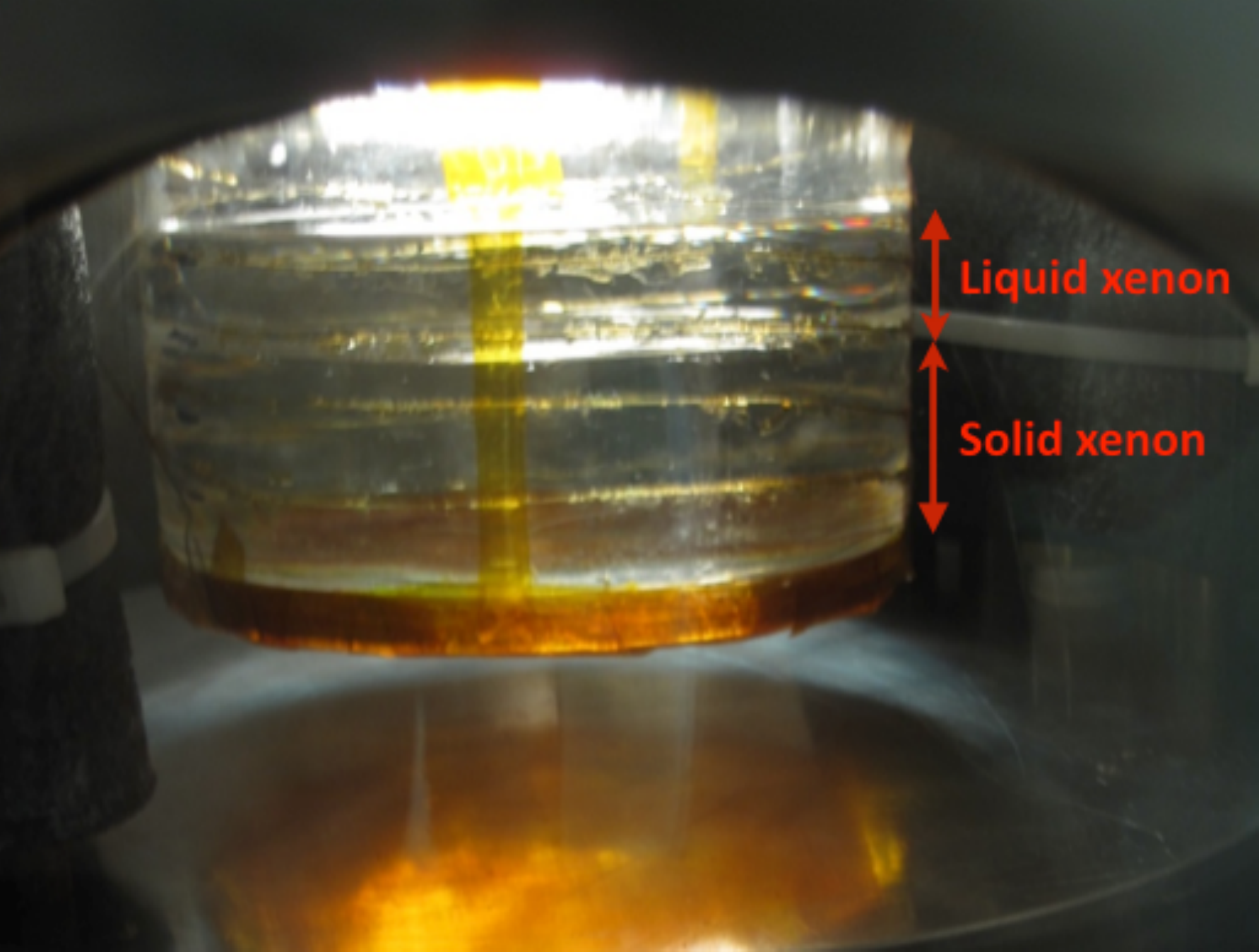}
\caption{Photographs of opaque solid xenon (left) and transparent solid xenon (right). The left photograph shows about 3\,cm of opaque solid xenon and 1\,cm of liquid xenon above the solid phase. The right photograph shows about 3\,cm high (about 650\,g) of transparent solid xenon and 1\,cm (about 189.4\,g) high  liquid xenon on top of the solid xenon.\label{fig:solidxenon}}
\end{center}
\end{figure}

\par Figure~\ref{fig:solidxenon} shows clear visual differences between frozen opaque xenon (left) and transparent solid xenon (right). Opaque xenon is produced when temperatures at the bottom of the glass chamber were reduced quickly ($\Delta T$/$\Delta t$<$\sim$3\,K/min). The transparent area above the opaque volume in the left photograph depicts xenon in the liquid phase. The right photograph in Figure~\ref{fig:solidxenon} shows about 3\,cm high solid xenon at the bottom of the chamber and about 1\,cm high liquid xenon on top of the solid xenon. In this case, the boundaries between the liquid phase and the solid phase can be seen by the reflection of light at the surface of the solid volume or refractive index changes at the boundary between liquid and solid. In the case of very well grown single phase solid xenon volumes, it is difficult to discriminate between the liquid phase and solid phase via visual inspection. 

Once the solid xenon is grown at the desired height we slowly raise the bottom temperature to the desired temperature, which is normally 157\,K (or 159\,K) while keeping the barrel temperature at 157\,K (or 159\,K). The largest volume of transparent solid xenon produced in the current solid xenon chamber was about 2\,kg at a temperature of 157\,K (see Figure~\ref{fig:sxetpc}, for example). The temperature of the large scale solid xenon can be further reduced with very slow control of temperature ($\Delta$T$\simeq$1\,K/day), however, we occasionally observed fuzzy opaque spots and structural defects in the solid volume at the temperature below 155\,K. Therefore, we limit our study down to 157\,K of solid xenon in the following scintillation and charge drift tests. 

\section{Scintillation readout}\label{sec:scintillation}
\par The scintillation properties of liquid xenon were studied in depth due to the advantage of a large photoelectric absorption cross-section for gamma rays, the ability to access the VUV scintillation light with quartz window PMTs, and a fast scintillation decay time below 20\,ns. The scintillation mechanism of the liquid xenon is understood as the decay of self-trapped excimer states which emits UV scintillation light (178\,nm)~\cite{jortner1965} and photons produced by ionization-recombination.

\begin{figure}[tb]
\begin{center}
\includegraphics[width=0.45\columnwidth]{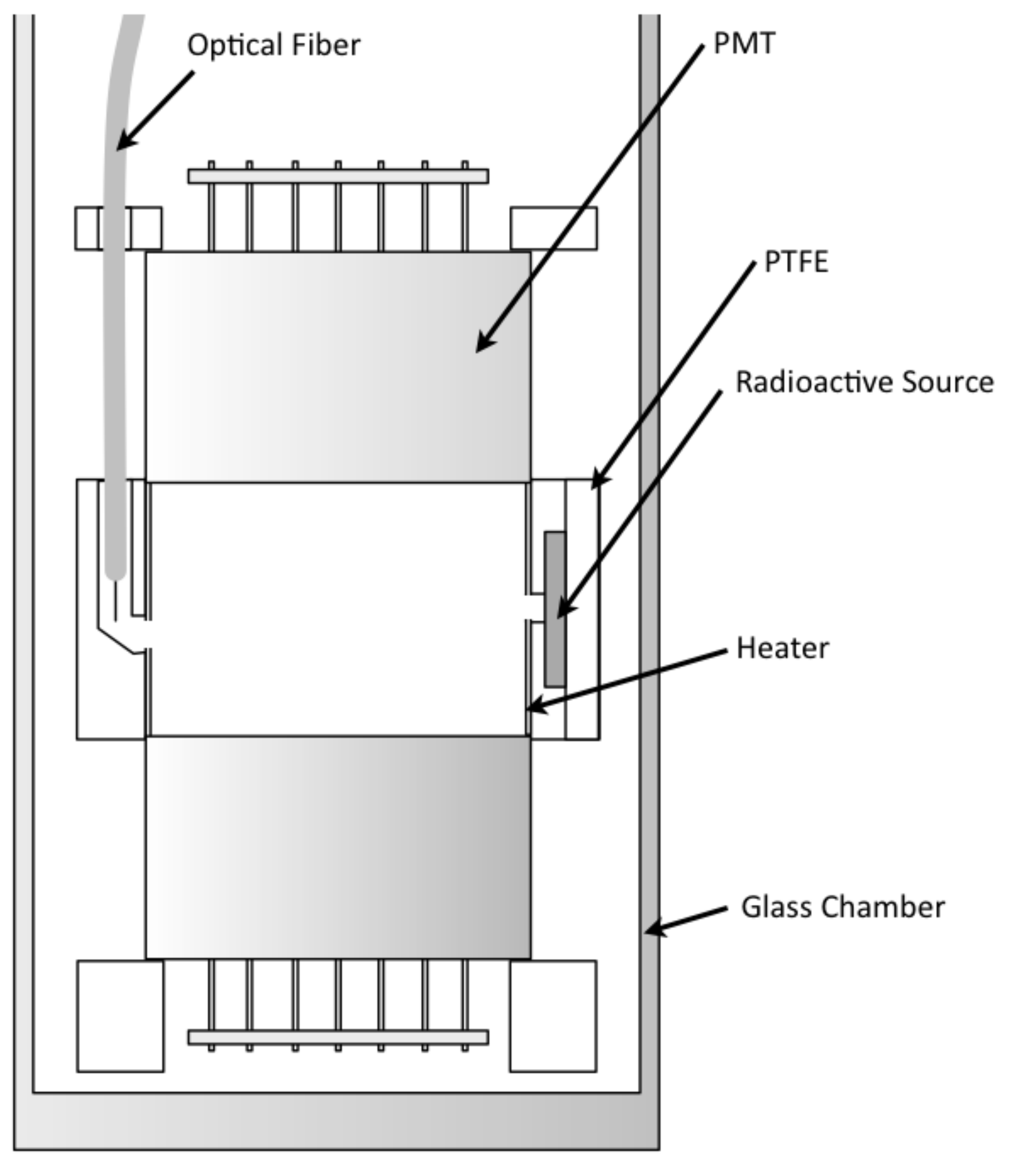}
\includegraphics[width=0.30\columnwidth]{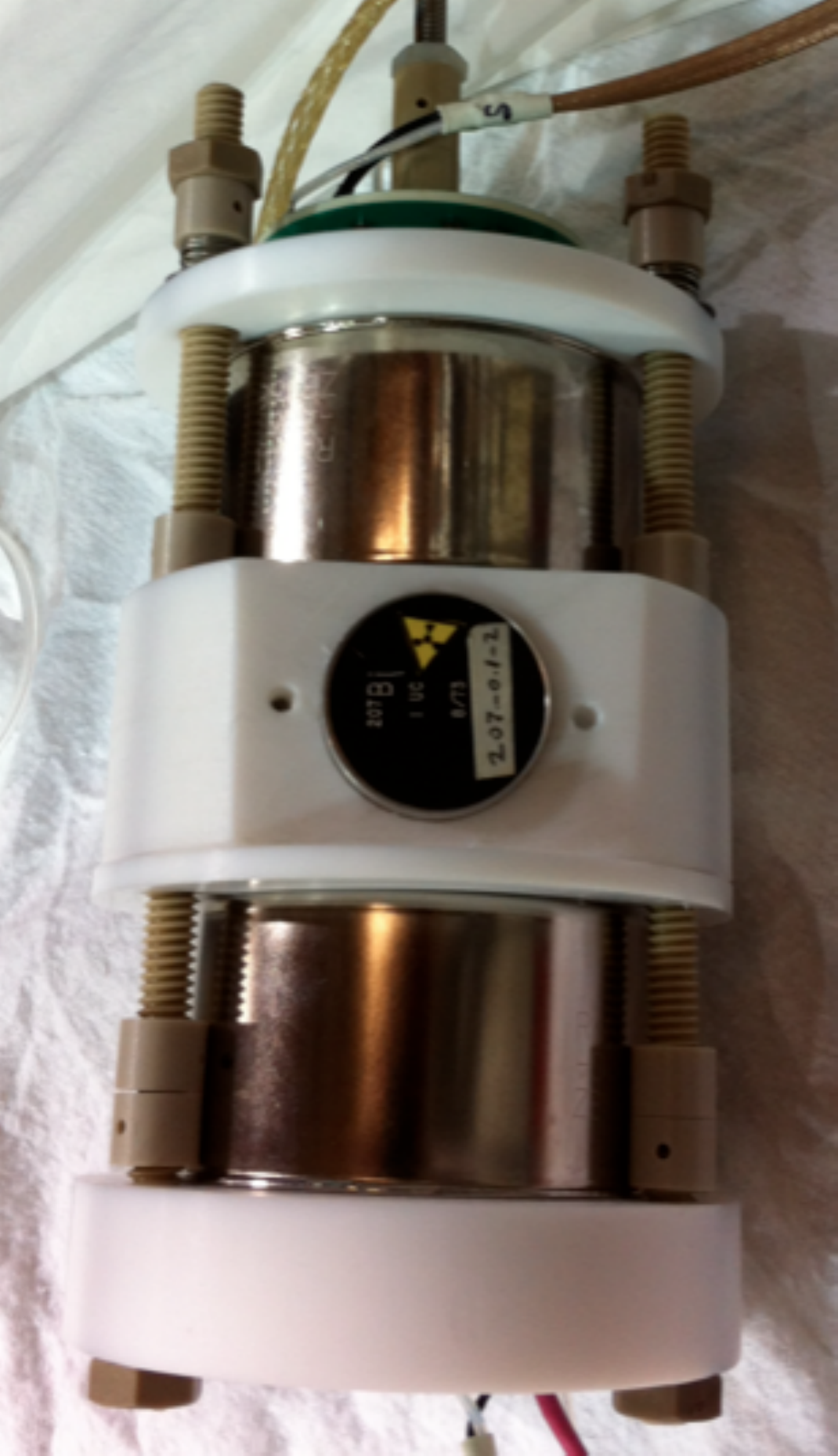}
\caption{A schematic drawing of the PMTs (left) and a picture of photo detector (left). The inner diameter of the detector is 5\,cm and height is about 3.8\,cm. See text for details. ~\label{fig:2pmt}}
\end{center}
\end{figure}

\par There are two major exciton decay modes in crystal noble elements; (1) a self-trapped exciton decay mode, and (2) a free exciton decay mode. The self-trapped exciton decay mode produces the well known $\sim$172\,nm range VUV photons ($\sim$7.2\,eV)~\cite{jortner1965}. The photon wavelength from the free exciton decay is temperature dependent. In the crystalline phase, the transport of free excitons may not be significantly interrupted by the dynamic phonon scatterings. These higher energy excitons can move freely quite a distance until they decay into the ground state. The wavelength from the free exciton decay is temperature dependent and can be as short as 148\,nm (8.4\,eV). A detailed spectroscopy study using thin layers ($\sim$1\,mm) of solid xenon at a temperature of 4.7\,K revealed these two clearly separate populations of the luminescence in the crystalline xenon~\cite{Varding1994, Reimand1999}.

\par In our study, we tested the scintillation light properties in large scale solid xenon under gamma radiation interactions using VUV PMTs. During the initial system test, we conducted studies dedicated to the optical coupling between the solid xenon and the PMT surface. One observation was that occasionally the optical coupling between solid xenon surface and PMT glass degrades if one simply solidifies xenon under normal pressure conditions (14.5 PSIA or lower). This is most likely due to the large density differences between liquid (2.95\,g/cc) and solid phase (3.41\,g/cc at 157\,K). We found the mechanical or optical contact of xenon onto the detector material substantially improved when the chamber was pressurized to 16 to 17\,PSIA. With this pressure, the xenon liquid above the solid volume slowly flows into the gaps between the structures and the optical coupling between the solid xenon-to-PMT can be made as good as liquid-to-PMT coupling.

\par Figure~\ref{fig:2pmt} shows a design and a photograph of the scintillation detector setup. The detector consists of two Hamamatsu R6041-406MOD VUV sensitive PMTs ($\sim$30\% Q.E at 178\,nm) enclosed by a Polytetrafluoroethylene (PTFE) tube. An optical fiber is guided into the center area of the detector and a Light Emitting Diode (LED) at the outside of the chamber is coupled to the optical fiber. The LED light source is used to calibrate the PMT gain and test the xenon transparency. A radioactive source holder is located at the side wall of the PTFE reflector. A PT-100 thermometer and a 32-gauge Nichrome heater wire is installed at the inner surface of the PTFE cylinder. This heater-thermometer loop controls the xenon temperature in the detector volume. 

\begin{figure}[tb]
\begin{center}
\includegraphics[width=6in]{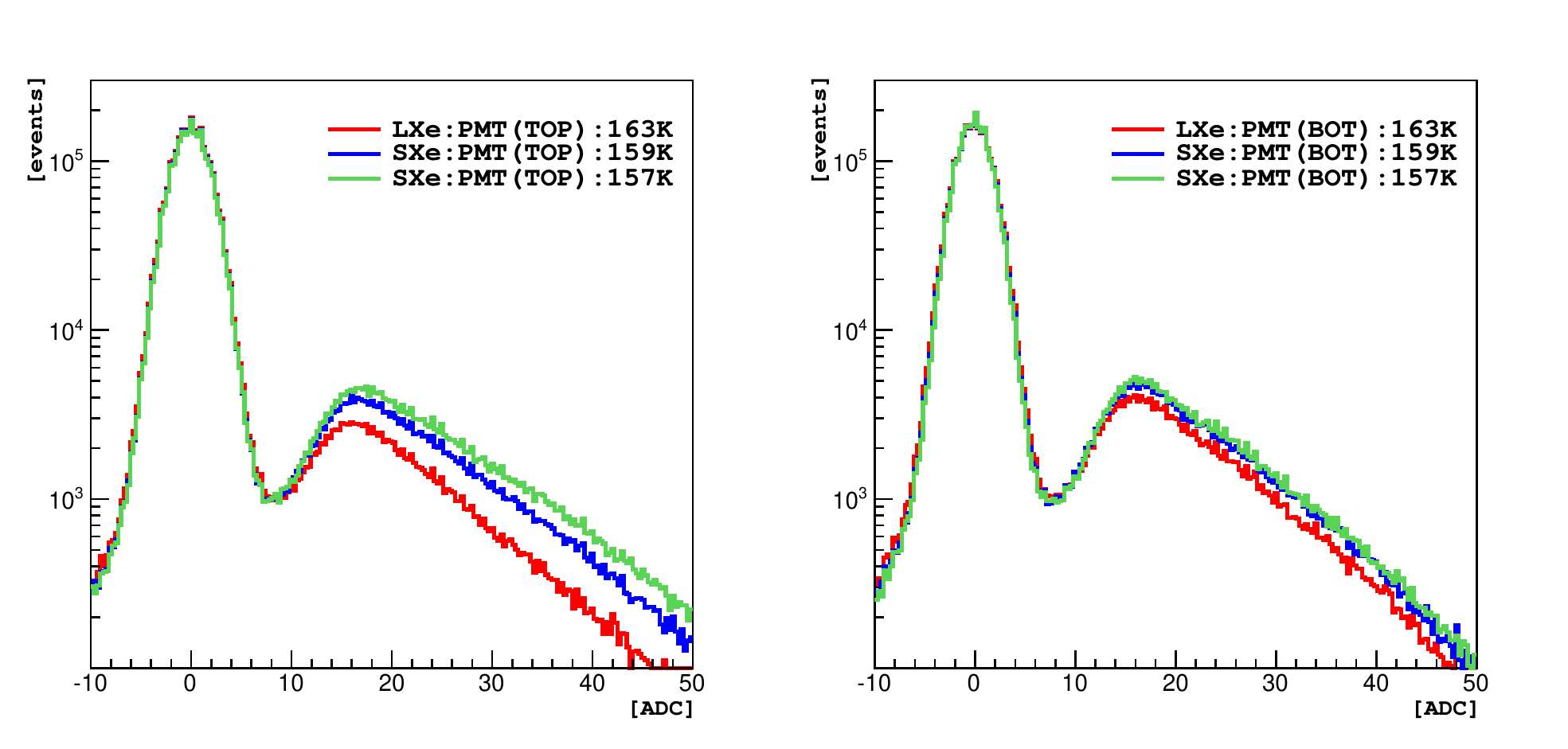}
\caption{Examples of SPE gains (high voltage of 830V) to the Hamamatsu PMT R6041-406MOD for top (left) and bottom (right) PMTs. The [ADC] on x-axis stands for the integral of the digitizer waveform after the baseline subtraction.~\label{fig:spe}}
\end{center}
\end{figure}

\par The scintillation light signals from the two PMTs are recorded using an 8-channel CAEN 1720B digitizer module at a sampling rate of 250\,MHz. A CAEN A2818 PCIE optical controller board is used to interface the digitizer and a desktop computer. The coincidence of the two PMT signals is used for the trigger. The trigger threshold is nominally set at a level of 0.3 SPE (single photo-electron). 

\par The integrated pulse area of the waveform scales the number of photo-electrons produced by the scintillation light. The gain of each PMT is calibrated by the SPE spectrum which is taken before and after each radioactive source run. The typical gain for the PMT at liquid xenon temperatures is about $10^6$ at an operating voltage of 830\,V. The SPE values of the corresponding source runs are used to correct the gains of the PMTs, therefore the systematic uncertainty of photon counting associated with the PMT gain is negligible. Figure~\ref{fig:spe} shows examples of SPE calibration of the PMT in both liquid and solid phase of xenon. The gain stability over temperature for the top (bottom) PMT is measured to be 2.5\% (1.7\%) in RMS. The gain difference between top and bottom PMTs is about 4.0\% at 157\,K, becoming 0.5\% at liquid xenon temperature (163\,K).  

\begin{figure}[tb]
\begin{center}
\includegraphics[width=5in]{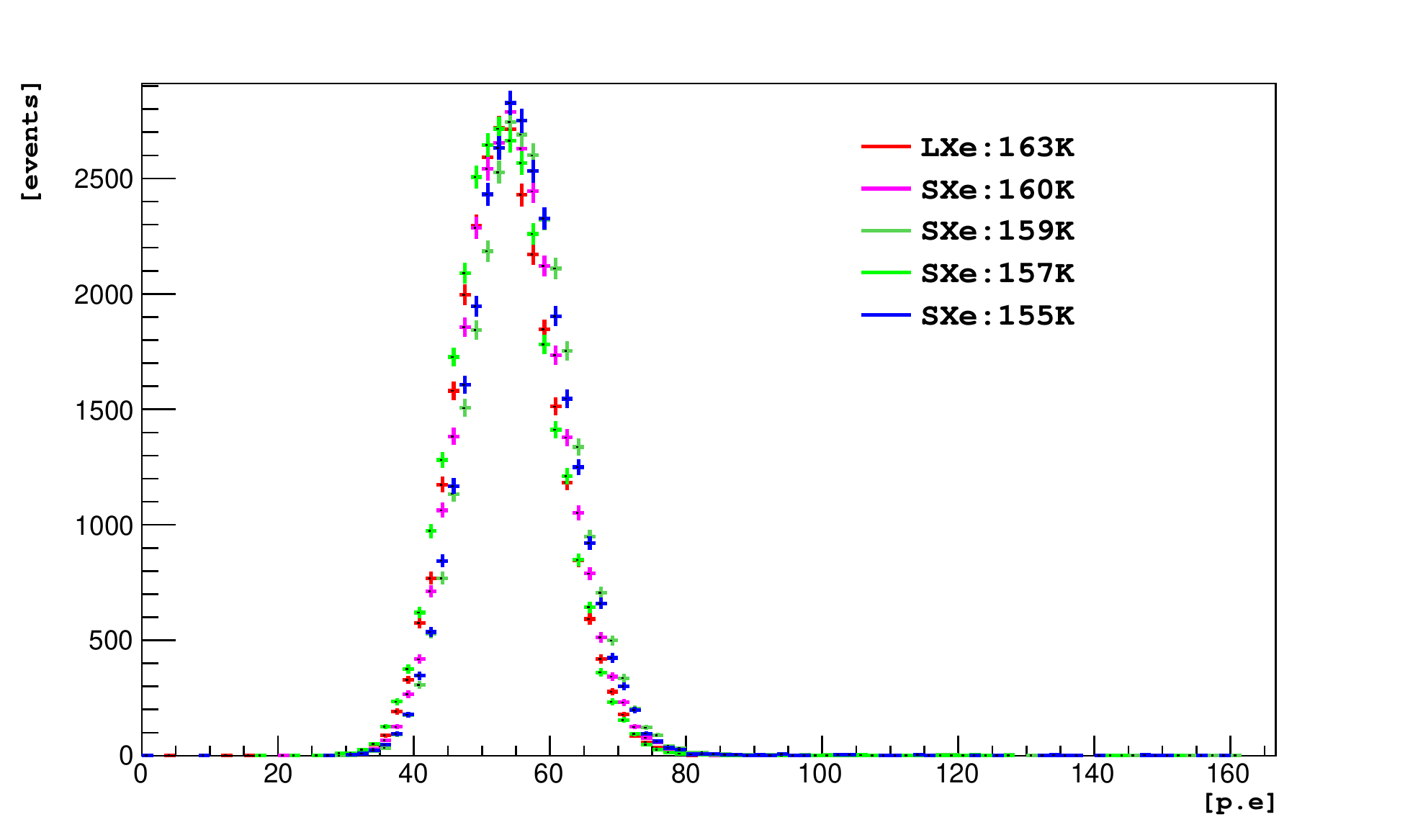}
\includegraphics[width=5in]{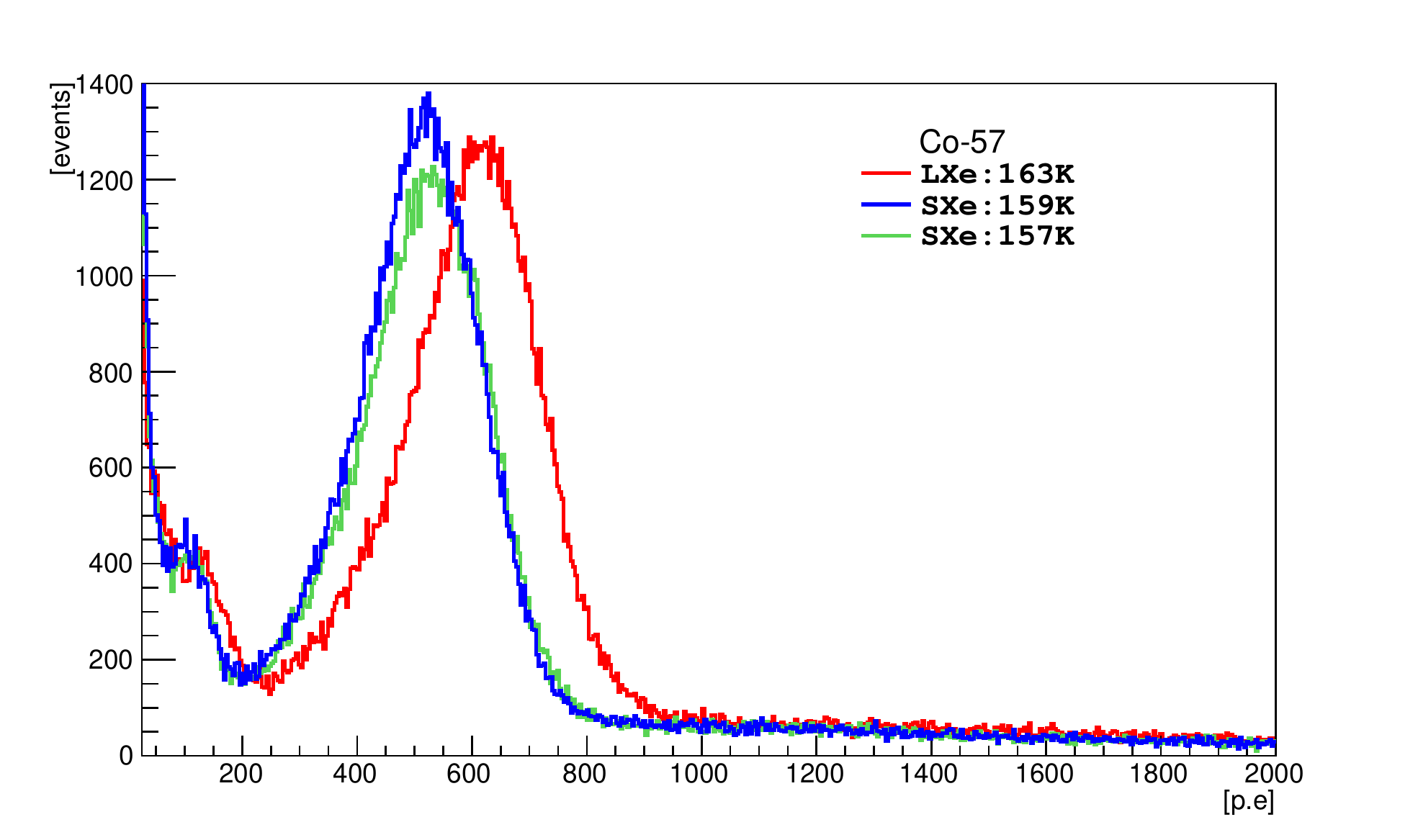}
\caption{Examples of multi-photon response of the detector in liquid and solid phases of xenon (top), and scintillation lights from gamma ray irradiation using $^{57}$Co source (bottom). ~\label{fig:mpe}}
\end{center}
\end{figure}

\par In order to prevent light leaks, the glass windows are covered with double layers of black shielding covers. The xenon chamber is no longer visually accessible for this measurement. To monitor the optical transparencies of the xenon volume, we carried out multi-photon occupancy tests using the LED source. Figure~\ref{fig:mpe} (top) shows the results from an example of this test in the liquid and solid xenon phases. We tuned the LED light source to obtain about fifty photo-electrons (sum of photo-electrons from two PMTs after the single p.e. corrections) in the liquid phase and compared that to solid xenon for various temperatures. The cooling process is handled carefully to maintain optical transparency. Using this method, we estimate the overall systematic uncertainty in photon counting to be about 4.2\% in the temperature range of 157\,K to 163\,K. We observed more photons in solid phase compared to that in the liquid phase while the LED pulser setup for multi-photon calibration remained unchanged. The variation of the LED intensity is estimated to be less than 1\%. However, the optical coupling between the edge of the optical fiber to the solid xenon may change compared to that in the liquid phase. Therefore, the multi-photon occupancy tests are used only to monitor the optical system changes during the operation, and not used to scale the photo-electron measurement of the scintillation light. 

\par We used a $^{57}$Co gamma source (14\,keV [9.54\%], 122\,keV [85.6\%], 136\,keV [10.6\%], and 692\,keV [0.02\%]) to understand the detector response. Figure~\ref{fig:mpe} (bottom) shows measurements of $^{57}$Co gamma calibration spectra at liquid and solid xenon temperatures. Each spectrum is normalized to constant detector exposure time. The results show consistent reduced light collection in the solid phase compared that in liquid phase. The reduced light collection in solid phase is most probably due to the changes in the refractive index. The LED multi-photon occupancy test shows no degradation of the optical transparency of the solid xenon compared to that in liquid phase.

\begin{table}[h!]
\begin{center}
\begin{tabular}{c| c c c c c c}
\hline 
                       & $N$=peak mean (p.e.)  & $\sigma$ (p.e.) & $\sigma$/$N$ & 2.35/$\sqrt{N}$ & FWHM & p.e. reduction \\
\hline
 163\,K &    603.6 $\pm$0.6   &  108.5 $\pm$0.6    &  0.180$\pm$0.001    & 0.096 &  0.45  & -     \\
 159\,K &    514.1 $\pm$0.5   &  \,\,98.1 $\pm$0.5 & 0.191$\pm$0.001 & 0.104  &  0.46  & 14.8\% \\
 157\,K &    523.7 $\pm$0.5   &  100.9 $\pm$0.5    &  0.193$\pm$0.001    & 0.103 &  0.49  & 13.2\% \\
\hline
\end{tabular}
\caption{Gaussian fit results for the combined 122 and 136 keV peak photo-electron distribution from $^{57}$Co for several temperatures. The table shows the mean and standard deviation obtained from the fit as well as calculations of the FHWM and the percent reduction in light output compared to a nominal temperature in liquid at 163\,K. }~\label{tab:co57fit}
\end{center}
\end{table}

\begin{figure}[tb]
\begin{center}
\includegraphics[width=5in]{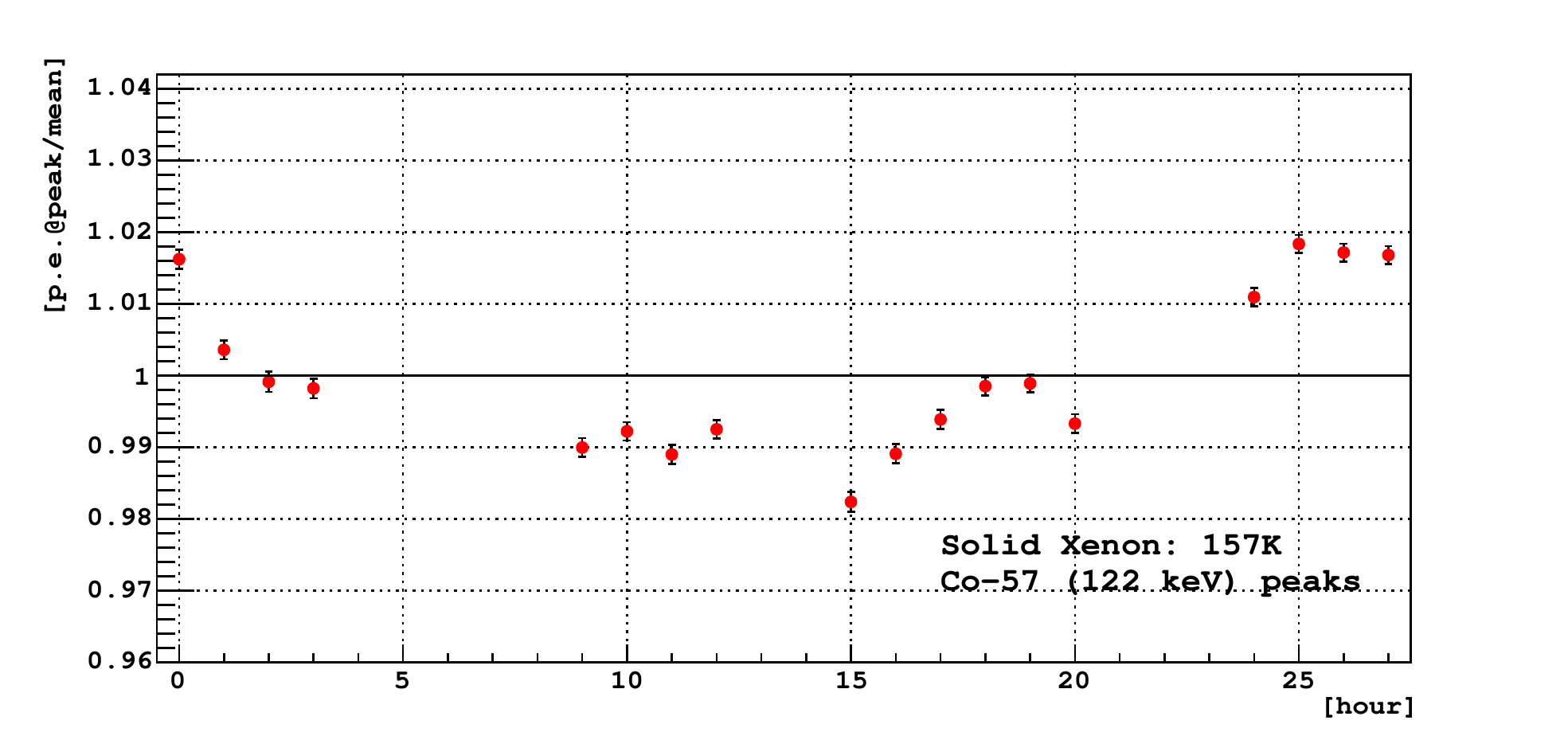}
\caption{Stability of the measured 122\,keV ($^{57}$Co) peaks over time at 157\,K. The vertical axis is defined by the number of photo-electrons divided by their average.~\label{fig:co57timevar}}
\end{center}
\end{figure}

We performed a Gaussian fit for the number of photo-electrons obtained from the combined 122 and 136\,keV gamma ray peaks assuming a first-order polynomial background. Table~\ref{tab:co57fit} shows the mean and standard deviation (sigma) obtained from the fit and the measured energy resolution at Full Width at Half Maximum (FWHM). The observed energy resolution (FWHM) at 122\,keV gamma peak in liquid is 0.45 broadening to 0.49 for solid xenon at 157\,K. The stability of the peak mean value at 157\,K over a day period is about $\pm$2\% and is shown in Figure~\ref{fig:co57timevar}. 

\section{Charge transport}\label{sec:chargetransport}
\par The properties of charge carriers such as drift velocity, diffusion coefficient, mean energy, and scattering cross-sections are important physics components in designing an ionization detector. These properties in thin layers of solid xenon have been measured in detail~\cite{Miller:1968zza, Gushchin1982}. One of the interesting observations in these measurements is the strong temperature dependence of the electron drift velocity in the condensed xenon. The drift speed of electrons at a given electric field in a thin solid layer is measured to be factors of two to five times faster than that in the liquid phase depending on the applied electric field and temperature. This can be understood as suppressed electron-phonon scattering in the solid phase due to the reduced energetic phonon populations in lower temperature media. 

\par The electro-negative contaminants are the dominant components that affect charge transportation. The microscopic defects, voids and structural deformation in a crystal can also affect the charge transport. Therefore, it is non-trivial to maintain the fast charge transport properties in a large scale solid. Due to the density changes between liquid to solid phase, the construction of solid xenon TPCs requires extra effort in order not to damage the TPC components during the solidifying process. The main focus of this test is, therefore, to measure the electron drift time in a kg scale of solid xenon using a conventional TPC device while keeping the optical transparency of TPC volume. 

\begin{figure}[tb]
\begin{center}
\includegraphics[width=3.8in]{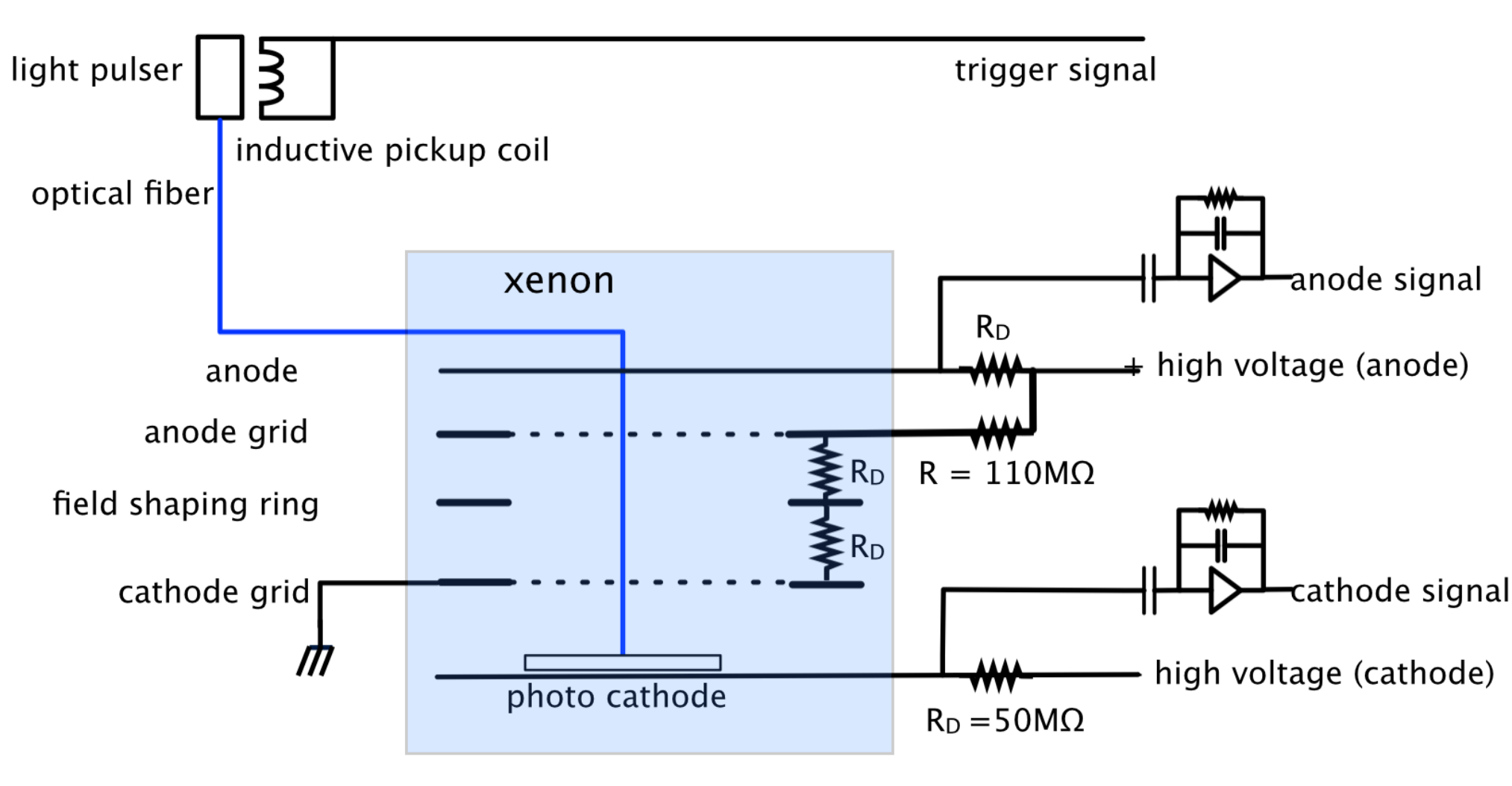}
\includegraphics[height=2.4in]{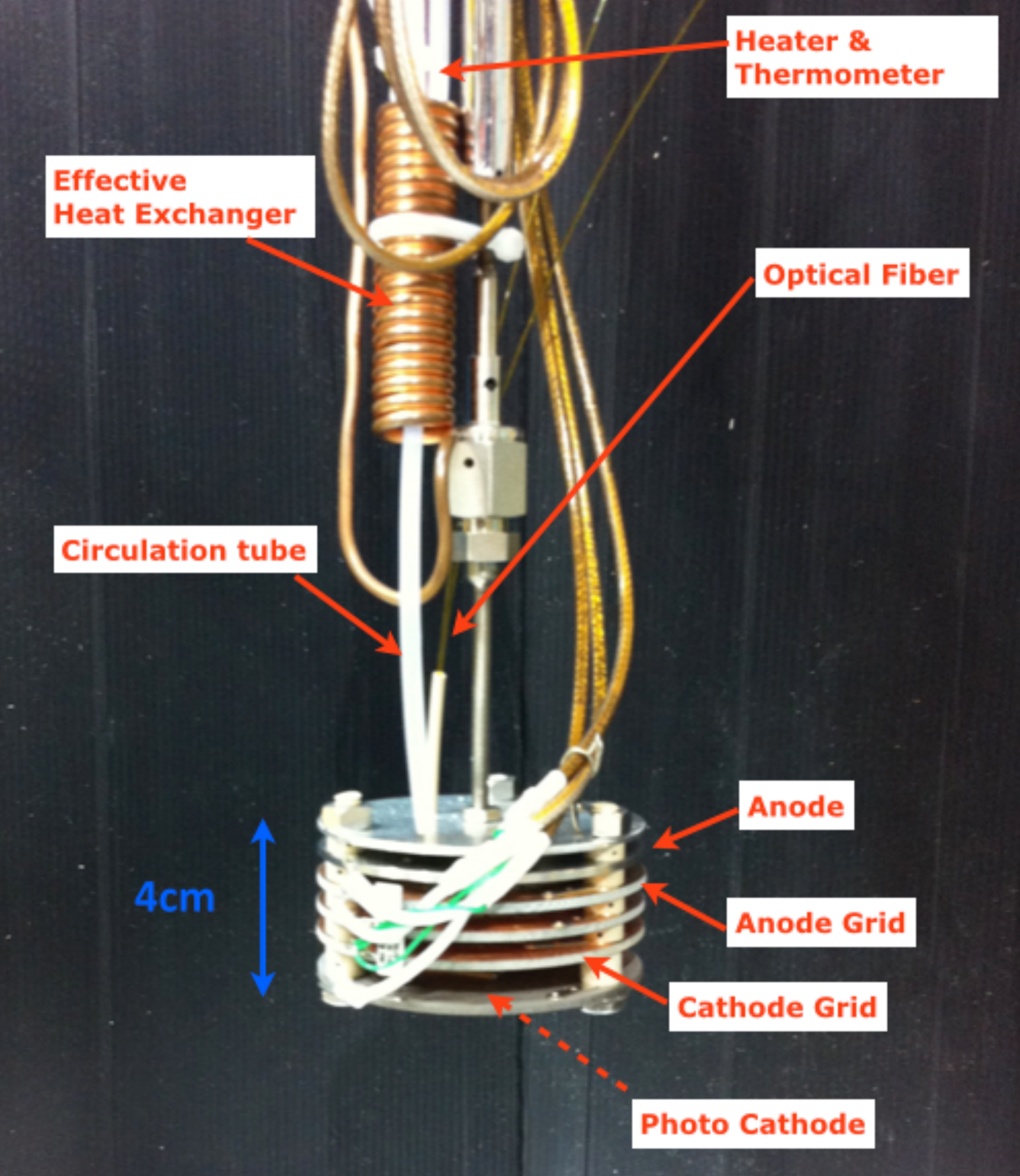}
\caption{A schematic of TPC (left) and a photograph of the TPC (right). \label{fig:tpc}}
\end{center}
\end{figure}

\par We adopted a charge drift electrode design reported in Reference~\cite{Carugno1990580}. This particular system employed a liquid argon purity monitor for liquid argon time projection chamber development at Fermilab~\cite{Adamowski:2014daa}. It consists of parallel, circular electrodes; a disk supporting a photocathode, two open wire grids, one anode and one cathode, and anode disks for the field shaping. The grid support rings are made of G-10 material, and the grids are made of electro-formed, gold-sheathed tungsten with a 2.0\,mm wire spacing and 25\,$\mu$m wire diameter. The geometrical transparency of the grid is 98.8\%. The anode grid and the field-shaping rings are connected to the cathode grid by an internal chain of 50\,M$\Omega$ resistors. The photocathode which is attached at the cathode disk is an aluminum plate coated with 50$\si{\angstrom}$ of titanium and 1000$\si{\angstrom}$ of gold. A xenon flash lamp, with a wide UV spectrum peaked at 225\,nm, was used as the
light source. The pulsed light is guided into the xenon chamber using two 0.6\,mm diameter core optical fibers. A trigger signal is produced using an inductive pickup coil on the power leads of the flash lamp. The photo-electrons from the photocathode drift to the cathode grid. After crossing the cathode grid, the electrons drift between the cathode grid and anode grids. Due to the electric field shielding by two grids, there is no current induced. After the electrons cross the anode grid, an electric current is induced at the anode. The signal is read out using charge amplifiers that have a 5\,pF integration capacitor and a 22\,M$\Omega$ resistor in parallel with the capacitor providing a typical time constant of 110\,$\mu$s.

\par Figure~\ref{fig:tpc} shows a schematic (left) of the short version of the TPC and a photograph of the TPC (right) installed in the xenon chamber. Two optical fibers are guided to the photocathode plate at the bottom of the TPC. The effective heat-exchanger and thermometer-heater system induces xenon circulation in the liquid phase. The length of the TPC from the cathode to the holding disk is 4\,cm. The actual drift distance of electrons from the photocathode surface to the anode is  3.0\,cm. The distance between photocathode surface to cathode-grid is 1.0\,cm (the thickness of photocathode is 0.2\,cm), cathode-grid to anode-grid is 1.5\,cm, and anode-grid to anode is 0.5\,cm. The thickness of each plate is about 0.2\,cm. In this short TPC, only one field shaping ring is installed in between the cathode-grid and anode-grid. A Tektronix 3034C Digital Phosphor Oscilloscope is used to readout the cathode and anode traces. The oscilloscope is connected to a PC through a GPIB communication protocol and LabVIEW software records the traces in a text file. 

\begin{figure}[tb]
\begin{center}
\includegraphics[width=4.5in]{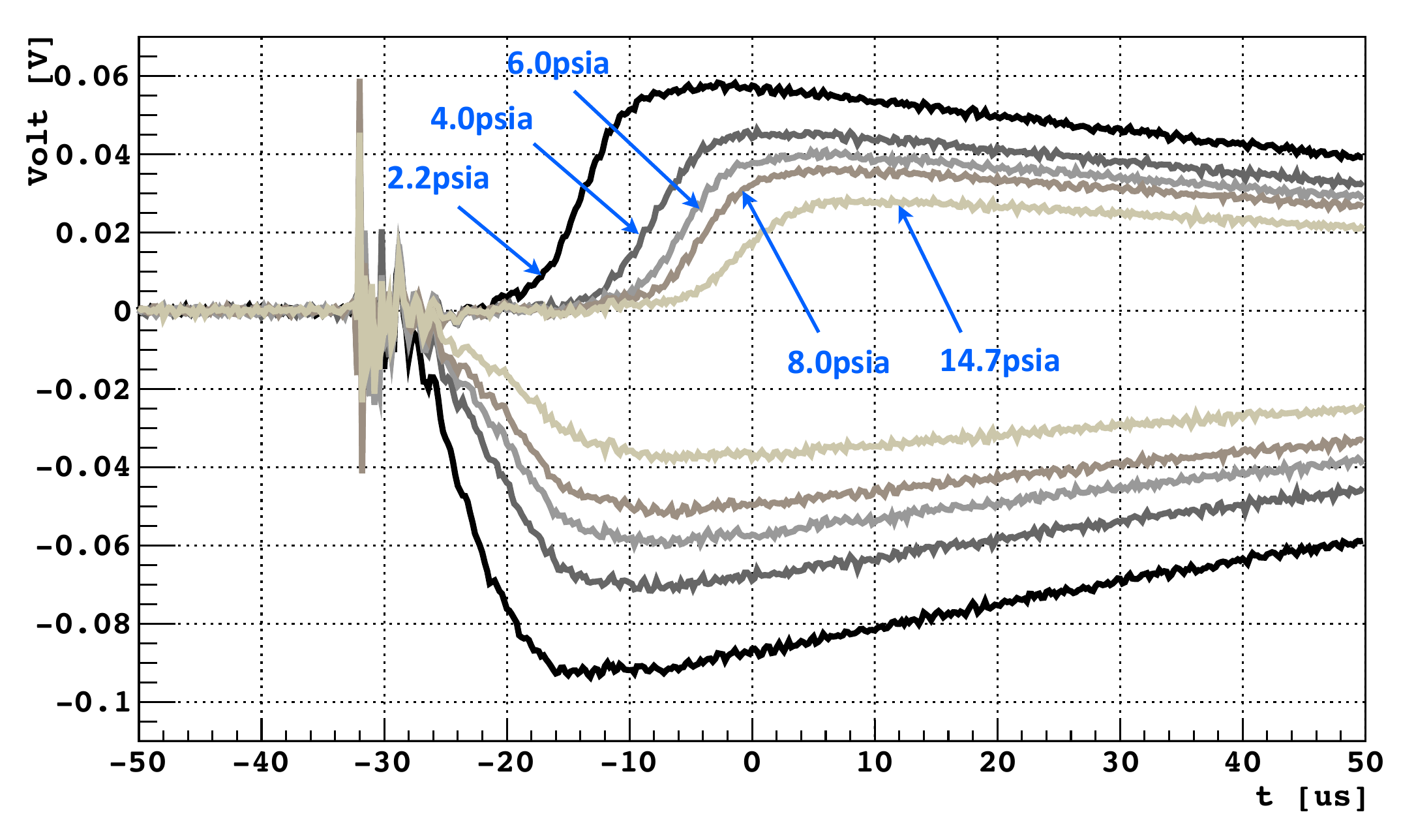}
\caption{TPC test under gas phase xenon for various pressures. The applied voltage between cathode and cathode grid is 300\,V/cm, cathode-grid to anode-grid is 200\,V/cm, and anode-grid to anode is 375\,V/cm. The traces above zero voltage are anode signals and the traces below zero are cathode signals.~\label{fig:gxetpc}}
\end{center}
\end{figure}

\par
\begin{figure}[t!]
\begin{center}
\includegraphics[width=2.8in]{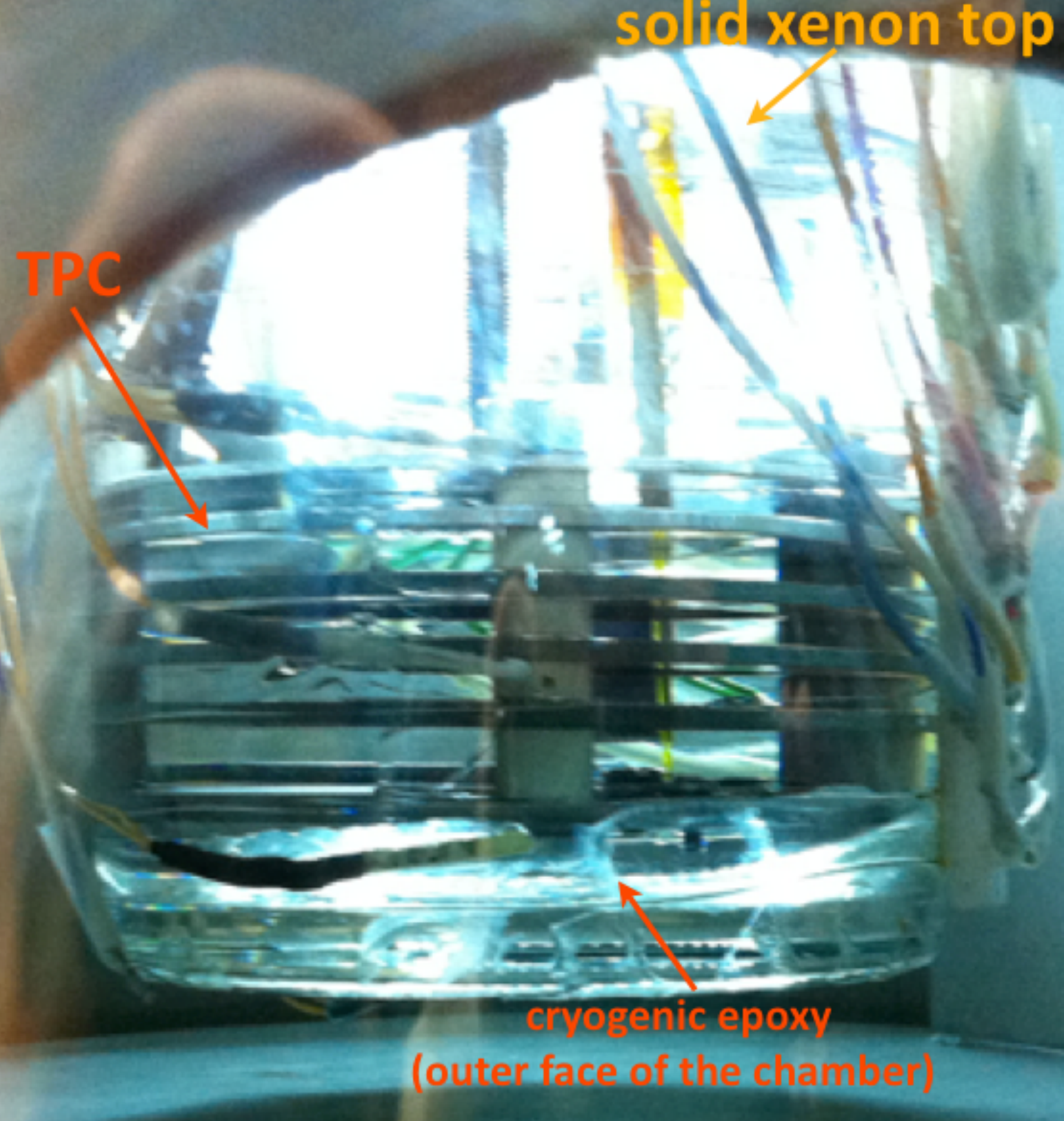}
\caption{The TPC installed in the xenon glass chamber. The TPC operation, such as pulsed photon emission through the optical fiber to the photo-cathode, can be viewed from the external window. In this particular sample, the amount of solid xenon in the chamber is about $\sim$2\,kg. A little opaque area near the surface of the glass bottom edge is cryogenic epoxy. In this particular sample, some birefringence can be seen at the inner glass wall near the grid rings.~\label{fig:sxetpc}}
\end{center}
\end{figure}

 We first made a TPC performance test by drifting electrons in xenon gas. Examples of such tests are shown in Figure~\ref{fig:gxetpc}. It is obvious that the electron drift time increases as a function of xenon gas pressure. Once the TPC is functioning properly in gas, the xenon is condensed to form the liquid phase. The xenon is continuously circulated through the getter until the charge signal is saturated. Depending on the initial xenon purity level, it takes a few days to achieve a large enough anode signal in the liquid phase. Even if the distance of the charge drift is relatively short (3\,cm), demonstration of the charge drift in over a kg scale of solid xenon is quite challenging, as the electro-negative purity level should be maintained over entire detector volume and the TPC component should not be damaged during the solidifying process. We measured the electron drift time in liquid first. Then the solid xenon was grown from the liquid (see Figure~\ref{fig:sxetpc}). The pulsed light emittance from the optical fiber to the photocathode is visually observable through the glass chambers. Good optical transparency was maintained during the electron drift measurement. 

\begin{table}[t!]
\begin{center}
\begin{tabular}{c  c  c | c  c  c  c}
\hline 
E$_c$       &   E$_{ag}$       & E$_{a}$  & $t_d$(L)        &  $t_d$(S)        &  $\tau$ (L)  & $\tau$ (S) \\

{\small [V/1.0\,cm]} &  {\small [V/1.5\,cm]} & {\small [V/0.5cm]} & {\small 164\,K~[$\mu$s]} & {\small 159\,K / 157\,K~[$\mu$s]}  & {\small 164\,K~[$\mu$s]} & {\small 159\,K / 157\,K~[$\mu$s] }  \\
\hline
 -1000 &   600  & 750  & 10.4$\pm$1.0 &  8.7$\pm$1.1 / 6.9$\pm$1.1   & 6.8 &  4.3 / 2.6   \\
 -1000 &   800  & 1000 & 10.9$\pm$0.6 &  8.6$\pm$1.0 / 6.6$\pm$1.1   & 5.9 &  3.5 / 3.5   \\
 -1000 &   1000 & 1250 & 11.2$\pm$0.6 &  7.5$\pm$0.6 / 6.5$\pm$0.6   & 5.4 &  4.6 / 3.1   \\
 -1000 &   1200 & 1500 & 11.1$\pm$0.6 &  8.2$\pm$0.6 / 6.9$\pm$0.6   & 5.5 &  3.7 / 3.4   \\
 -1000 &   1400 & 1750 & 11.1$\pm$0.6 &  7.8$\pm$0.6 / 7.2$\pm$0.6   & 5.4 &  4.1 / 2.4   \\
 -1000 &   1600 & 2000 & 10.9$\pm$0.6 &  7.9$\pm$0.6 / 6.5$\pm$0.6   & 5.4 &  3.6 / 3.2   \\
 -1000 &   1800 & 2250 & 10.9$\pm$0.6 &  7.8$\pm$0.6 / 6.3$\pm$0.6   & 5.1 &  3.7 / 3.6   \\
 -1000 &   2000 & 2500 & 10.8$\pm$0.6 &  7.5$\pm$0.6 / 6.7$\pm$0.6   & 5.0 &  4.0 / 3.0   \\
\hline  
\end{tabular}
\caption{Measured electron drift time for various electric fields and temperatures. The total electron drift distance for this measurement is 3.0\,cm. The fitting uncertainty of the $\tau$(L) and $\tau$(S) measurements are all less than 0.1$\mu$s, which are omitted in the table for simplicity.}\label{tab:drifttime}
\end{center}
\end{table}

\begin{figure}[!htp]
\begin{center}
\includegraphics[width=5.2in]{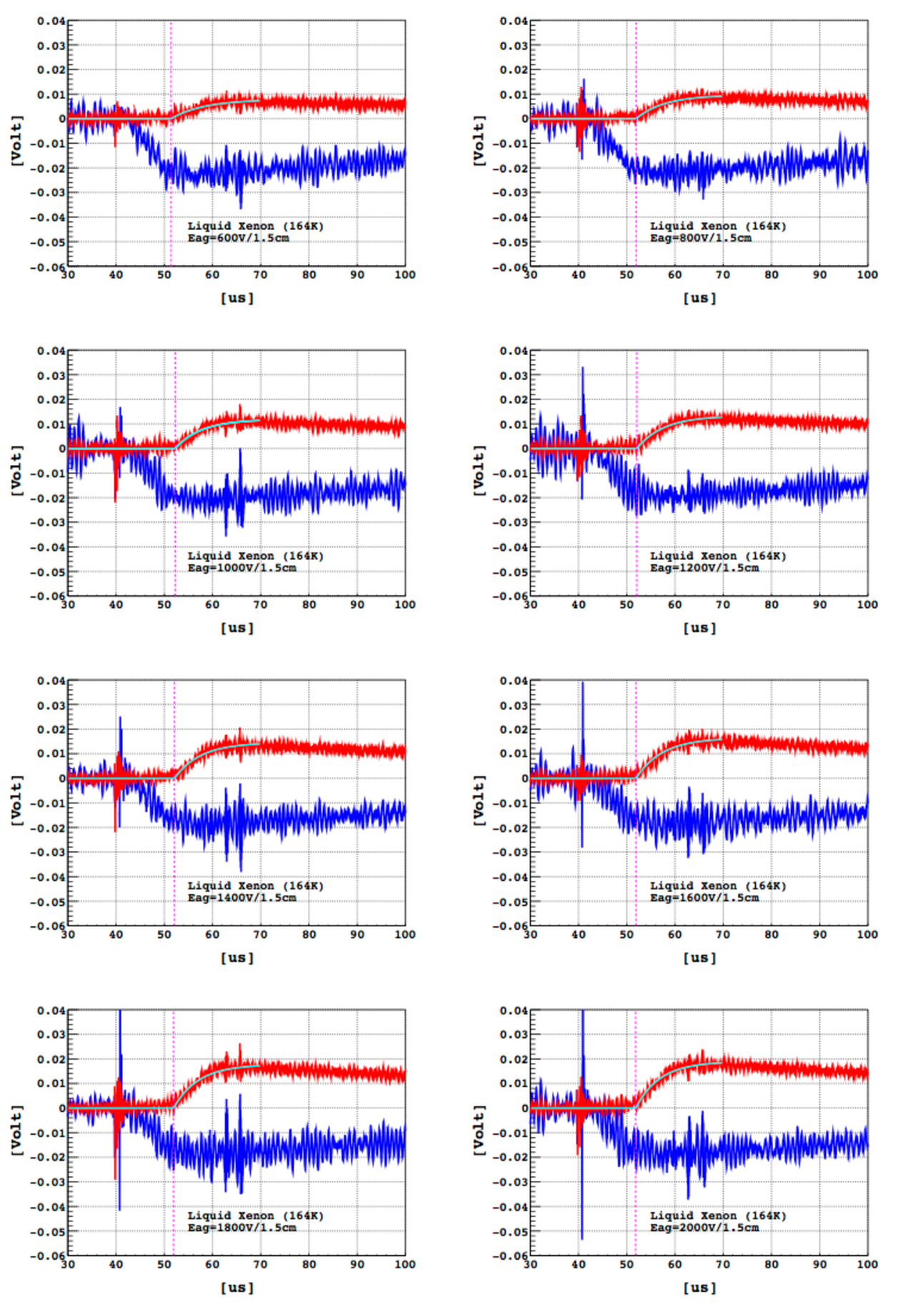}
\caption{Measurements of electron drift time for different electric fields in liquid phase of xenon (164\,K). The red curves are the anode signals and the blue curves are the cathode signals. The green curves are the fit functions. The dotted magenta lines indicate the anode start times ($t_0$ in the fit function, see text for details).}\label{fig:edriftliquid}
\end{center}
\end{figure}
\begin{figure*}[!htp]
\begin{center}
\includegraphics[width=5.2in]{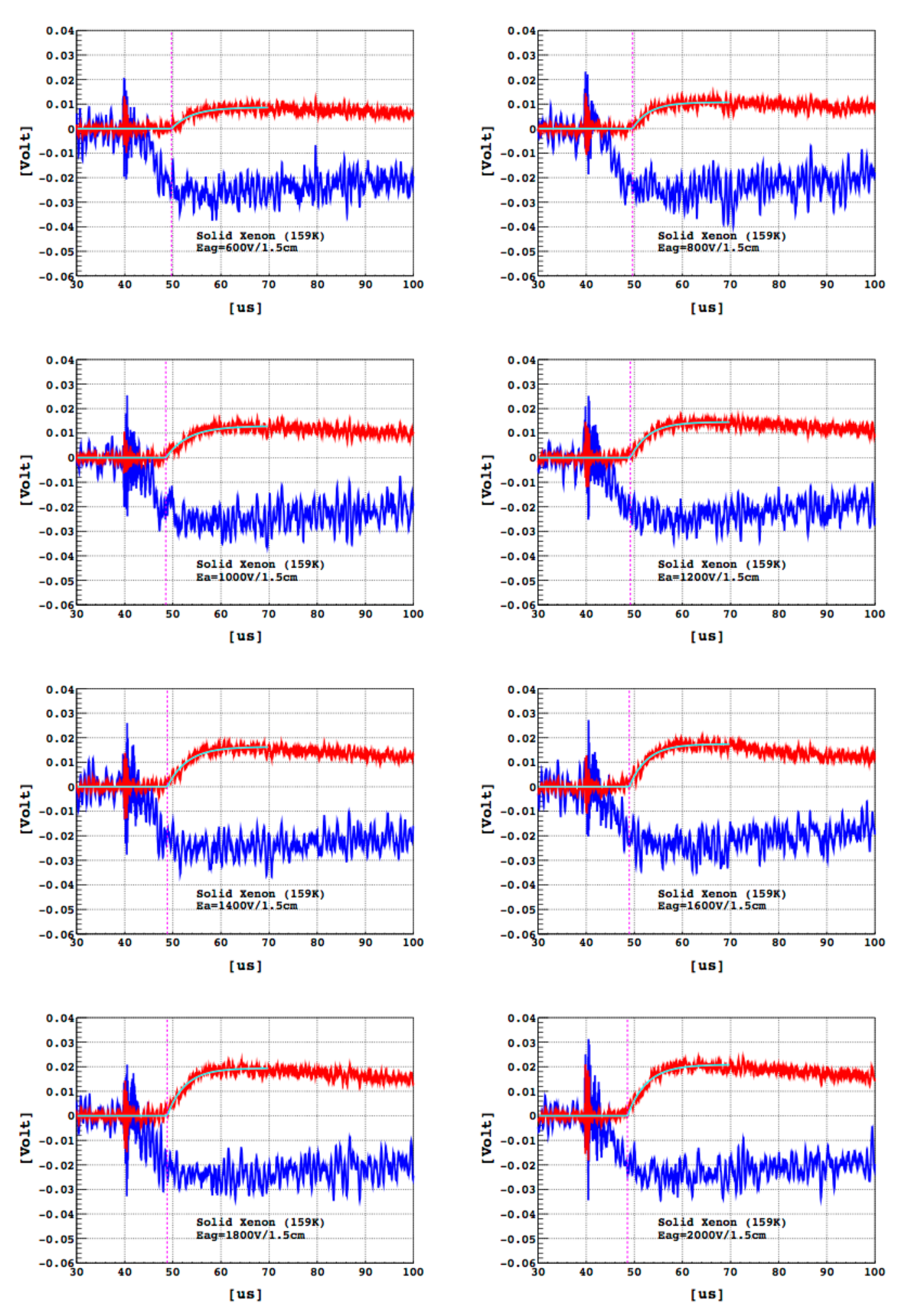}
\caption{Measurements of electron drift time for different electric fields in solid phase of xenon (159\,K). The red curves are the anode signals and the blue curves are the cathode signals. The green curves are the fit functions. The dotted magenta lines indicate the anode start times.}\label{fig:edriftsolid1}
\end{center}
\end{figure*}
\begin{figure}[!htp]
\begin{center}
\includegraphics[width=5.2in]{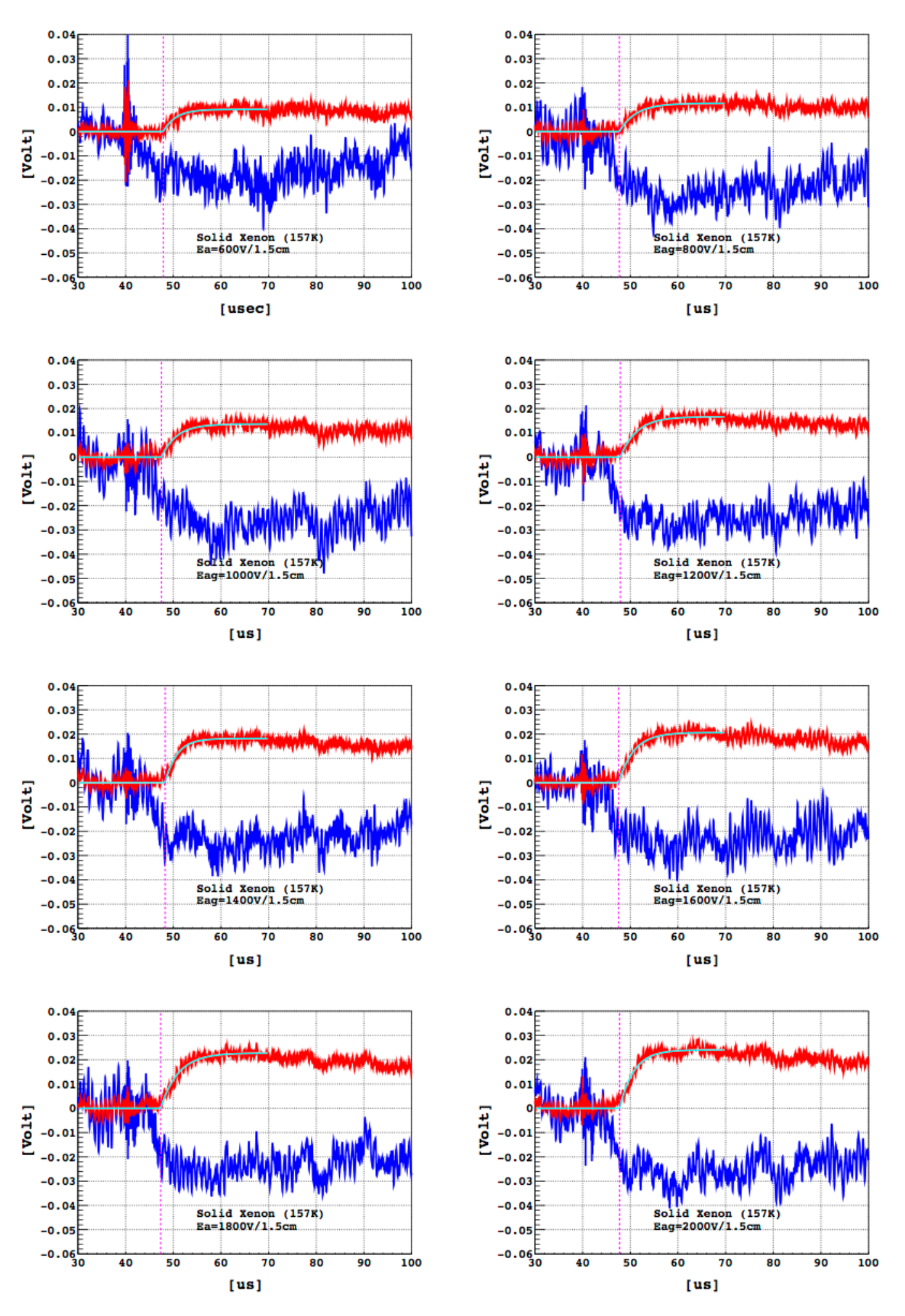}
\caption{Measurements of electron drift time for different electric fields in solid phase of xenon (157\,K). The red curves are the anode signals and the blue curves are the cathode signals. The green curves are the fit functions. The dotted magenta lines indicate the anode start times.}\label{fig:edriftsolid2}
\end{center}
\end{figure}

Table~\ref{tab:drifttime} shows results of the measurements. Figures~\ref{fig:edriftliquid},~\ref{fig:edriftsolid1}, and~\ref{fig:edriftsolid2} show signals from the anode and cathode in liquid and solid xenon at different temperatures and at several applied electric fields. A total of 64 traces are averaged to cancel the white noise. The residual background noise in the trace has been subtracted using the trace of a zero-field measurement. The sharp peaks at around 41\,$\mu s$ are electric glitch from the light pulser, which are not completely removed after the noise subtraction. Some remaining noise in the traces come from the heater power source, which switches AC to DC, and noise pickup from the high-speed vacuum turbo. 

\par To find the anode start time ($t_0$), we fit a function to the anode trace, given as $f(t) = A (1-exp((t-t_0)/\tau)$, where $A$, $\tau$ and $t_0$ are free parameters. To directly compare the drift time of electrons between the liquid phase and solid phase, we used the trigger signal from the inductive pickup coil as a start time of the electron creation ($t_i=41\mu$s$\pm$0.5$\mu$s). The electron drift time is defined by $t_d = t_0-t_i$. Table~\ref{tab:drifttime} shows the summary of these measurements. The electric field $E_{ag}$ is chosen to maximize the electron transparency providing significant amplification at the anode signal, while varying the applied voltage at E$_a$. 

\par The uncertainty of the electron drift time measurement is from the fitting error of $t_0$, and error of the $t_i$ offset ($\pm 0.5\,\mu$s) which is dominated by the electric noise near $t_i$. The large measurement error for E$_{ag}$=600\,V/1.5\,cm and 800\,V/1.5\,cm is due to the small amplitudes of the anode signals in the low electric field measurement. The errors are determined through repeated measurement at the same electric field. The uncertainty in the rise time fit parameter ($\tau$) is less than 0.1\,$\mu$s, however the overall fluctuation of the $\tau$(L) is 2.0$\mu$s, and $\tau$(S) is 1.1$\mu$s in RMS. The electron drift speed in solid phase xenon at 157\,K (at 159\,K) is about 1.6 (1.4) times faster and the rise time of the anode signal is about 1.8 (1.4) times faster than in those of the liquid phase at the given electric fields in this test.

\section{Discussion}\label{sec:discussion}
\par The R\&D effort reported here is focused on particle detector applications of solid xenon with three goals: (1) proof of scaleability of optically transparent solid xenon, (2) measuring the scintillation light, and (3) demonstrating electron drift in a large scale solid xenon and measuring the drift time. 

\vspace{0.5cm}
\noindent{\bf Scalability}\par
We demonstrated the scalability of transparent solid xenon above a kg scale and found an optimal, efficient method. A modified {\it Bridgeman's technique} is suited to consistently reproduce optically transparent solid xenon. The solid xenon production is automatically controlled using heater and thermometer loops in a gas-cooled glass chamber system and the technique can be extended to develop a normal cryogenic chamber system which has no direct optical access. It took more than two days starting from liquid phase to grow a 9\,cm in diameter and 9\,cm high cylinder of solid xenon. The transparency can be kept for an extended period of time under stable temperatures and pressures. The longest time that we kept the solid xenon in the chamber was about 10-days. Our system is designed to maximize visual access to the xenon volume while keeping the glass chamber in a safe condition. However, it is obviously not the best design for a uniform thermal configuration. For example, a cold-bath method, where the xenon chamber slowly sinks into a temperature-calibrated cold liquid bath, would be a very attractive solution in order to build a larger scale detector. We found vapor deposition method is still good option for a low temperature and a thin-layer (less than a cm scale) of solid xenon detector development. We often made about a 5\,mm thick transparent solid xenon bulk at the temperature of 77\,K via vapor deposition method. 

\vspace{0.5cm}
\noindent{\bf Scintillation readout}
\par We observed a systematic reduction in the number of photoelectrons in the solid phase compared to that in the liquid phase. Based on the multi-photon tests, we estimate a systematic uncertainty of about 4\% in photon collection between the solid and liquid phases. This is attributed to changes in the critical angle at the boundary between the optical fiber and the xenon bulk liquid or solid. Table~\ref{tab:refindex} summarizes the indices of refraction of optical components and critical angle of optical boundaries for xenon scintillation wavelength (172\,nm) and the mean wavelength (500\,nm) of the LED lights. The measured reduction of scintillation light in the solid phase may also be understood by changes in the xenon refractive index, giving rise to changes in total internal reflection at the xenon-PMT glass boundary. The critical angle at this boundary is smaller for solid xenon. Thus, compared to liquid xenon, more photons will undergo total internal reflection in solid, then be subject to more reflections off the teflon walls. 
\begin{table}[t!]
\begin{center}
\begin{tabular}{l | c  |  c }
\hline 
Index of refraction       &   $n$ ($\lambda=172$\,nm (S), 178\,nm (L))  &   $n$ ($\lambda=500$\,nm) \\
\hline
Optical fiber (core)      &    -                     &    1.62   \\
PMT synthetic silica window    &    1.61(172\,nm), 1.59 (178\,nm)                  &   1.54    \\
Liquid xenon (164\,K)     &    1.70                  &   1.39    \\
Solid xenon (157\,K, 159\,K)  & 1.90                 &   1.45    \\
\hline 
\hline 
Critical angle ($\theta_c = \arcsin(n_2/n_1)$) & $\theta_c$ ($\lambda=178$\,nm (S), 172\,nm (L))  & $\theta_c$ ($\lambda=500$\,nm)\\
\hline
Optical fiber ($n_1$) $\rightarrow$ liquid xenon ($n_2$) &  -  &  59.1$^{\circ}$\\
Optical fiber ($n_1$)$\rightarrow$ solid xenon ($n_2$)&  -  & 63.5$^{\circ}$\\
Liquid xenon  ($n_1$)$\rightarrow$ PMT window ($n_2$)& 69.3$^{\circ}$  & $>90^{\circ}$\\
Solid xenon ($n_1$)$\rightarrow$ PMT window ($n_2$)&  57.9$^{\circ}$  & $>90^{\circ}$\\
\hline
\end{tabular}
\caption{Indices of refraction of the optical system and critical angles of the total internal refraction ($\theta_c$) for each optical boundary. The indices of refraction for different xenon phases (S:Solid and L:Liquid) for different scintillation wavelength are from the references~\cite{Sinnock1980, Hitachi2005}. }~\label{tab:refindex}
\end{center}
\end{table}
\par
\begin{figure}[t!]
\begin{center}
\includegraphics[width=6in]{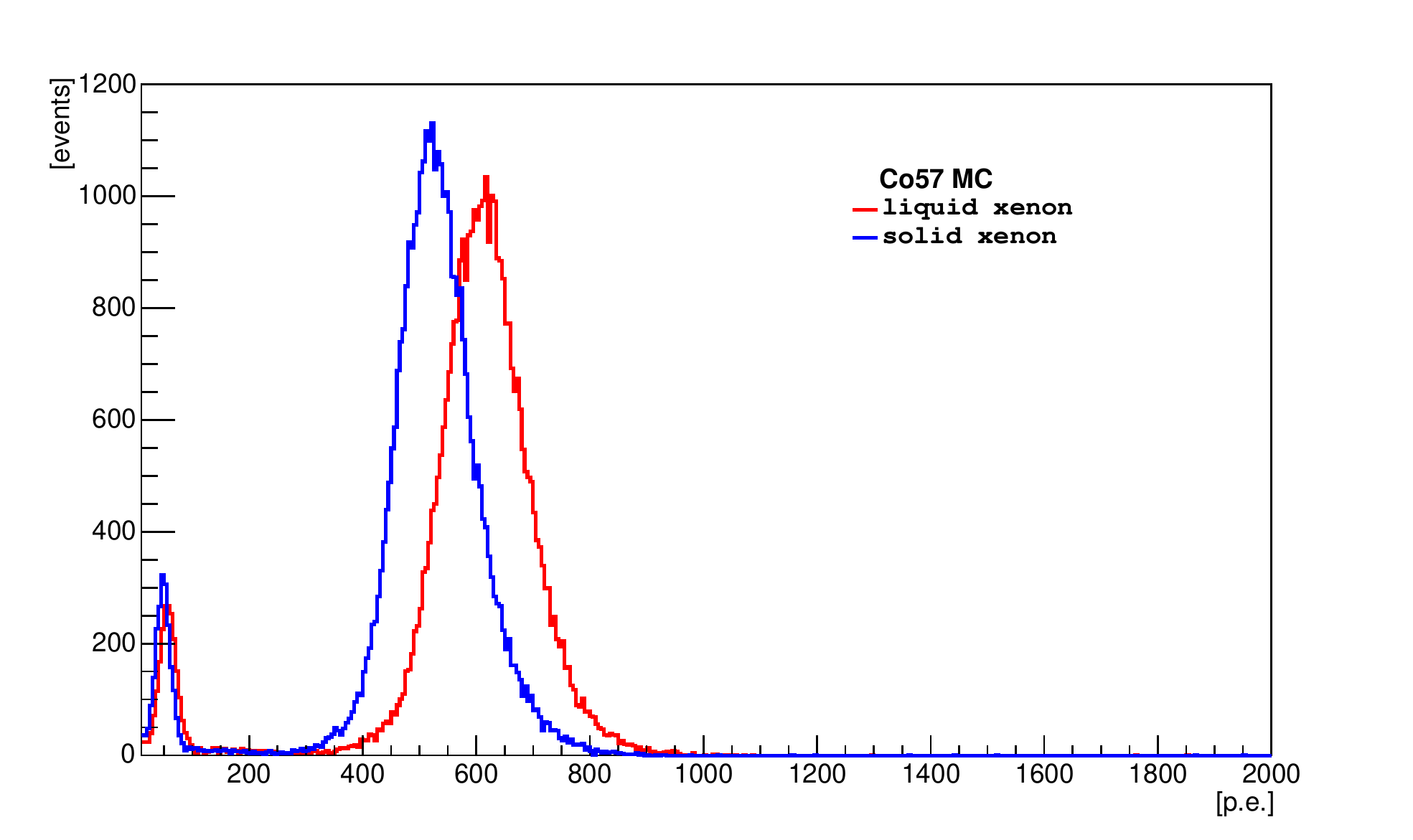}
\caption{The simulated energy distribution of $^{57}$Co radioactive decays for liquid and solid xenon.}\label{fig:comc}
\end{center}
\end{figure}

 We carried out Monte Carlo simulations in order to understand if these $\theta_c$ corrections can explain the amount of reduced scintillation photon collection in the solid phase. Events from the decay of $^{57}$Co are simulated using a dedicated event generator~\cite{Ponkratenko:2000um}. Particles from the simulated source are then transported in the xenon bulk volume using the GEANT4~\cite{Agostinelli:2002hh} (version 4.9.6 patch 2) framework coupled to a scintillation light production module for xenon~\cite{Szydagis:2011tk}. The model accounts for all relevant xenon properties including the non-linear scintillation light yield as a function of incident gamma ray energy. We assume the same scintillation light yield model in liquid and solid phase simulations and account for the differences in the xenon emission spectrum in each phase. However, we used different scintillation wavelength in liquid (178\,nm) and solid (172\,nm) which effectively changes the energy dependent W$_s$ value (the average energy required per one scintillation) in solid by 3.4\% in the model. In fact, the W$_s$ in solid has some ambiguity in references~\cite{RGSv2,Obodovski79}. The model also accounts for all relevant wavelength-dependent material properties. We assume a xenon bulk absorption length of 100 cm and Rayleigh scattering length of about 30\,cm~\cite{Baldini:2004ph}. We account for the wavelength-dependent quantum efficiency of the PMTs. We found the major uncertainty of the light collection efficiency is from the PTFE reflectivity. For the PTFE housing, we assume a reflection coefficient of 0.57 at the xenon scintillation wavelengths. Figure~\ref{fig:comc} shows the results of the simulation. This simulation suggests a 14.2\% reduction in the number of photo-electrons expected in the solid phase compared to liquid, consistent with measured values (see Table~\ref{tab:co57fit}). The measured energy resolution at 122\,keV in liquid xenon is about 0.45 (FWHM). The energy resolution of this system depends on photo-electron statistical fluctuations, fluctuations of electron-ion recombination due to escaped electrons, the non-proportionality of the scintillation light yield, single p.e. response of the PMTs and effects from the detector geometry~\cite{Ni:2006zp}. According to the simulation results, the energy resolution contributed by the photo-electron statistics, detector geometry and optical system is about 0.272 (FWHM). The ratio of the measured width of the single p.e distribution to its mean is measured to be 0.3\,(=s) in sigma, which contributes to the energy resolution about 0.092 in FWHM (=2.35/$\sqrt{(1+s^2) \cdot N}$). The rest of the fluctuation may not be solely attributed by the fluctuations of the electron-ion recombination but some unknown detector systematics, which deserves further study with improved detector setup. 

\par Another possible interpretation of reduced light collection in solid xenon would be the partial creation of free-exciton (FE) decay modes in the xenon crystal, which can produce shorter wavelength (below $\sim$160\,nm) scintillation light. These shorter wavelength photons are non-transparent to the quartz windows, causing reduced photon collection~\cite{Kink1987}. However, given the quantitative consistency of the Monte Carlo simulation results with the observed light reduction, the FE decay effect may not be clearly visible in the presented test, if we assume the energy resolution is a separate systematic issue. Further study with a setup using short wavelength photo-sensing devices (such as CsI photocathodes~\cite{Aprile1994129, Buzulutskov2008}) or wavelength shifting photo-devices with reduced optical systematics might clarify the observed light reduction issue. 

\vspace{0.5cm}
\noindent{\bf Charge drift}
\par We measured electron drift time in large scale optically transparent solid xenon using a conventional TPC. There are two findings in these results. First, the drift velocity of electrons was about 1.6 (1.4) times faster in the solid phase at 157\,K (159\,K) compared to that in liquid phase at 164\,K at the given electric field configuration. Second, the rise time (the fit parameter $\tau$) of the anode charge signal is about 1.8 (1.4) times faster in solid phase at 157\,K (159\,K) compared to that in liquid phase at 164\,K at the given electric field. The results are qualitatively consistent with the measurements of using a thin layer of solid xenon in reference~\cite{Gushchin1982}. The large uncertainties in drift time measurements mainly came from the electric noise at the beginning of the pulses. A TPC with additional field shaping rings between the cathode and anode grids could reduce the relative uncertainties of the drift time measurement. 
 
\vspace{0.3cm}
\par Considering the demonstrated optical transparency of a large scale solid xenon, it is plausible to build a dual-phase (solid-gas) xenon scintillation detector with high-voltage drift field setup. In this setup the first scintillation signal (S1) in solid xenon and the secondary luminescence signal (S2) enhanced by electrons in gas phase xenon can be used for accurate measurement of the electron drift time. The dual-phase design can also be extended to use a multi-anode PMTs or Silicon Photo Multipliers in gas phase to measure the particle tracks~\cite{Bolozdynya1977} by drifting and extracting electrons from the solid phase.

\section{Summary}\label{sec:summary}
\par The research and development efforts towards employing solid xenon as a particle detector was presented. Using a three-chamber vessel with automated liquid nitrogen cooling conditions, we demonstrated the scalability of optically transparent solid xenon above a kilogram scale. Scintillation light from gamma ray irradiation in the solid xenon was detected using two VUV sensitive PMTs. We  observed reduced numbers of photo-electrons in the solid phase compared to that of the liquid phase at given gamma energies. This effect is most probably due to the optical system changes between liquid and solid xenon. Further study using more advanced photo-sensing systems with reduced optical systematic uncertainties might be able to address this issue in the future. We measured the charge drift speed in solid xenon using a conventional TPC system. The electron drift speed in solid phase xenon at 157\,K (159\,K) is about 1.6 (1.4) times faster and the rise time of anode signal is 1.8 (1.4) times faster than in those of liquid phase at the given sets of electric field configuration in this test. We hope this initial study motivates future advanced detector development, such as fast response TPC design using ultra-cold solid xenon, charge tracking detectors using pixelated sensors on solid xenon, and low energy bolometric phonon readout detectors at sub-milliKelvin temperatures.

\acknowledgments
We are very grateful to M.~Miyajima, J.~White, and A.~Bolozdnya for the initial discussions of the solid xenon particle detector and sharing their ideas. We thank R.~Barger, D.~Butler, R.~Davis, A.~Lathrop, L.~Harbacek, K.~Hardin, C.~Kendziora, B~.Miner, K.~Taheri,  M.~Rushmann, E.~Skup, M.~Sarychev and J.~Vorin at Fermilab for their tireless hard work to provide us the experimental setup with highest standard. We also thank V.~Anjur, A.~Anton and B.~Loer for their participation of the system setup. This work supported by the Department Of Energy Advanced Detector R\&D funding.

\bibliographystyle{h-physrev3}
\bibliography{sxref}{}
\end{document}